\documentclass[12pt, a4paper]{article}
\pdfoutput=1
 \usepackage[font=small,format=plain,labelfont=bf,up,textfont=normal,up,justification=justified,singlelinecheck=false]{caption}

\usepackage[a4paper, left=2cm,right=2cm]{geometry}
\usepackage[colorlinks=true,linkcolor=black,citecolor=teal,urlcolor=MidnightBlue,filecolor=black]{hyperref}
\usepackage{amsfonts}
\usepackage{amsmath}
\usepackage{setspace}
\usepackage[dvipsnames]{xcolor}
\definecolor{SchoolColor}{rgb}{0.6471, 0.1098, 0.1882} 

\usepackage[utf8,applemac]{inputenc}
\usepackage{tensor}
\usepackage{cite}
\usepackage{tikz}
\usepackage{graphicx}
\usepackage{bm} 

\usepackage[utf8,applemac]{inputenc}
\usepackage{tensor}
\usepackage{cite}
\usepackage{tikz}
\usepackage{graphicx,subfigure}
\usepackage[font=small, font+=it, format=hang, margin={1cm, 1cm}]{caption}
\graphicspath{{figure/}}
\usepackage{array}
\usepackage{booktabs} 
\usepackage{multirow}

\usepackage{dcolumn}
\usepackage{bm}

\numberwithin{equation}{section}
\newcommand{\bea}{\begin{eqnarray}}
\newcommand{\eea}{\end{eqnarray}}
\newcommand{\be}{\begin{equation}}
\newcommand{\ee}{\end{equation}}
\def\nn{\nonumber}

\newcommand{\beqs}{\begin{eqnarray}}
\newcommand{\eeqs}{\end{eqnarray}}
\numberwithin{equation}{section}

\begin{document}
\begin{titlepage}

\begin{flushright}\vspace{-3cm}
{\small
\today }\end{flushright}
\vspace{0.5cm}
\begin{center}
	{{ \LARGE{\bf{Logarithmic behaviour of \vspace{6pt}\\
					connected correlation function in CFT}}}} \vspace{5mm}

	\centerline{\large{\bf{Jiang Long\footnote{e-mail:
					 longjiang@hust.edu.cn}}}}
	\vspace{2mm}
	\normalsize
	\bigskip\medskip

	\textit{School of Physics, Huazhong University of Science and Technology, \\Wuhan, Hubei 430074, China
	}
	
	\vspace{25mm}
	
	\begin{abstract}
		\noindent
		{We study $(m)$-type connected correlation functions of OPE blocks with respect to one spatial region in two dimensional conformal field theory. We find logarithmic divergence for these correlation functions. We justify the logarithmic behaviour from three different approaches: massless free scalar theory,  Selberg integral and conformal block. Cutoff independent coefficients are obtained from analytic continuation of conformal blocks. A UV/IR relation has been found in connected correlation functions. We could derive a formal ``first law of thermodynamics'' for a subsystem using deformed reduced density matrix. Area law of connected correlation function in higher dimensions is also discussed briefly.}  	\end{abstract}
	

\end{center}

\end{titlepage}
\tableofcontents

\section{Introduction}
Divergent problem in continues quantum field theory is not a catastrophe. Usually, it can be cured by introducing an energy scale (UV cutoff) in the theory. This is still true in computing entanglement entropy in conformal field theory (CFT)\cite{Bombelli:1986rw,Srednicki:1993im,Callan:1994py,Araki:1976zv}. One famous result is that one interval R\'enyi entropy in CFT$_2$ presents logarithmic behaviour and univeral \cite{Calabrese:2004eu}. In higher dimensions, it also diverges and obeys area law \cite{Eisert:2008ur}. R\'enyi entropy, roughly speaking, is a generator of connected correlation functions of modular Hamiltonians
\be
\langle H_A^m\rangle_c. \label{ham}
\ee Therefore, each connected correlator of modular Hamiltonians \eqref{ham} shows logarithmic behaviour in CFT$_2$. 
 
On the other hand, motivated from holographic duality of modular Hamiltonian \cite{Jafferis:2014lza, Jafferis:2015del}, we showed that connected correlation functions of modular Hamiltonians 
\be
\langle H_A^mH_B^n\rangle_c,\quad m,n\ge1
\ee 
are finite \cite{Long:2019fay}. For CFT$_2$, they are also universal due to conformal Ward identity \cite{Belavin:1984vu}. More generally, the author claimed that $(p,m-p)$-type connected correlator 
\be
\langle Q_A[\mathcal{O}_1]\cdots Q_A[\mathcal{O}_p]Q_B[\mathcal{O}_{p+1}]\cdots Q_B[\mathcal{O}_{m}]\rangle_c, \quad p\ge1,m\ge2\label{mntype}
\ee 
is also finite \cite{Long:2019pcv}, where $Q$s are OPE blocks \cite{Ferrara:1971vh,Ferrara:1972cq, Czech:2016xec}. Unlike $(m,n)$-type correlator, $(m)$-type correlator
\be
\langle Q_A[\mathcal{O}_1]\cdots Q_A[\mathcal{O}_m]\rangle_c
\ee 
 is divergent in general. There is no way to remove  divergent terms in these correlators.  Therefore, one nature problem is to understand the divergent behaviour of $(m)$-type connected correlation functions.

In this paper, we show that $(m)$-type correlator may present logarithmic behaviour like \eqref{ham} in CFT$_2$
\be
\langle Q_A[\mathcal{O}_1]\cdots Q_A[\mathcal{O}_m]\rangle_c=C[h_i]\log\frac{L}{\epsilon}
\ee
where $L$ is the length of interval and $\epsilon$ is UV cutoff.
Therefore one could still extract cutoff independent information from coefficient $C[h_i]$. We extract $C[h_i]$ using a method  of  regularization  developed by \cite{Long:2019fay}. 
We could validate these coefficients using three approaches. Firstly, we prove the logarithmic behaviour in massless free scalar theory. Secondly, we obtain the same divergent behaviour by analytic continuation of Selberg integrals \cite{Selberg:1944}.  We also find that the logarithmic behaviour of $(m)$-type correlator can be obtained from analytic continuation of $(p,m-p)$-type correlator, which is equivalent to take the limit $B\to A$ of \eqref{mntype}. Since $(m-1,1)$-type correlator is claimed to be conformal block \cite{Long:2019pcv}, 
\be
\langle Q_A[\mathcal{O}_1]\cdots Q_A[\mathcal{O}_{m-1}]Q_B[\mathcal{O}_m]\rangle_c=D[h_i]G_{h_m}(\eta),
\ee
we could prove the logarithmic behaviour by analytic continuation of conformal block for general $(m)$-type correlators.  As a consequence of this observation, we obtain a UV/IR relation between coefficients $C[h_i]$ and $D[h_i]$
\be
C[h_i]=\frac{2(-1)^{h_m}\Gamma(2h_m)}{\Gamma(h_m)^2}D[h_i],
\ee 
where $C[h_i]$ characterizes the divergent behaviour when $B$ coincides $A$  (UV) and $D[h_i]$ characterizes the leading order behaviour when $B$ and $A$ are far away (IR).
Our result shows that deformed reduced density matrix 
\be
\rho_A=e^{-W},\quad W=\sum_i \alpha_i Q_A[\mathcal{O}_i]
\ee 
encodes rich information of CFT$_2$, its vacuum expectation value 
\be
\log\langle\rho_A\rangle=F(\alpha_i)\log\frac{L}{\epsilon}\label{2dlog}
\ee 
also presents logarithmic divergence. We could derive a formal ``first law of thermodynamics'' associated deformed reduced density matrix for any subsystem
\be
d\mathcal{E}_A=\Omega_A+TdS_A+\mu \mathcal{Q}_A, 
\ee 
where $\mathcal{E}_A, \Omega_A, T, S_A,\mu$ and $\mathcal{Q}_A$ are ``energy'',``grand potential'',``temperature'',``Gibbs entropy'',
``chemical potential'' and ``charge'' of the subsystem in parallel to statistical mechanics.

We generalize the observation of analytic continuation of conformal block to higher dimensions. Interestingly, we find an area law 
\bea
\langle Q_A[\mathcal{O}]^m]\rangle_c\sim \left\{\begin{aligned}&\gamma\ \frac{\mathcal{A}}{\epsilon^{d-2}}+p_{d-4}(\frac{R}{\epsilon})^{d-4}+\cdots+q_1\log\frac{R}{\epsilon}+q_0+\cdots,\quad&\text{d odd},\\&\gamma\ \frac{\mathcal{A}}{\epsilon^{d-2}}+p_{d-4}(\frac{R}{\epsilon})^{d-4}+\cdots+q_2\log^2\frac{R}{\epsilon}+q_1\log\frac{R}{\epsilon}+q_0+\cdots,\quad&\text{d even}.\end{aligned}\right.
\eea 
The logarithmic behaviour differs from usual R\'enyi entanglement entropy for general primary operator. 
 When operator $\mathcal{O}$ is a conserved current $\mathcal{J}$ whose dimension $\Delta$ and spin $J$ satisfy
\be
\Delta=J+d-2,\quad J\ge 1,
\ee 
the area law becomes 
\bea
\langle Q_A[\mathcal{J}]^m]\rangle_c\sim \left\{\begin{aligned}&\gamma\ \frac{\mathcal{A}}{\epsilon^{d-2}}+p_{d-4}(\frac{R}{\epsilon})^{d-4}+\cdots+q_0+\cdots,\quad&\text{d odd},\\&\gamma\ \frac{\mathcal{A}}{\epsilon^{d-2}}+p_{d-4}(\frac{R}{\epsilon})^{d-4}+\cdots+q_1\log\frac{R}{\epsilon}+q_0+\cdots,\quad&\text{d even}.\end{aligned}\right.
\eea
It shows similar structure as R\'enyi entanglement entropy.

The structure of this paper is as follows. In Section 2 we will review known results on connected correlation function and indicate that $(m)$-type correlators  obey logarithmic law in CFT$_2$. We will check this logarithmic behaviour up to $m\le 4$ in Section 3. After that, we will work on massless free scalar theory to support our observation. In Section 5, analytic continuation of Selberg integral has been used to justify the logarithmic behaviour. We establish the relation between $(p,m-p)$-type correlator and $(m)$-type correlator in Section 6. We find more logarithmic behaviour by analytic continuation of conformal block to another region in the following section. In Section 8, we study the expectation value of deformed reduced density matrix. Motivated by the similarity between deformed reduced density matrix and density matrix of thermodynamic system, we establish a formal ``first law of thermodynamics'' for any subsystem. In section 9, we discuss the area law of connected correlation function in higher dimensions.  We conclude with a discussion of future directions for this program in the last section. Technical details on conformal blocks, integrals and analytic continuation are collected in four appendices.

\section{Review}
Modular Hamiltonian \cite{Haag:1992}, which is the logarithm of reduced density matrix, plays a central role in the evolution of a subsystem. The subsystem we are interested in is one interval $A$ of a CFT$_2$. The total system is in Minkowski spacetime and the state is in vacuum. We use $(t,z)$ to denote spacetime coordinates. The domain of dependence of $A$ is $D(A)$. The endpoints of interval A are $x_1=-1,x_2=1$. In other words, the radius of $A$ is $1$ and the center is at origin. Therefore the length of interval $A$ is 
\be
L=2.
\ee  Modular Hamiltonian for one interval is \cite{Casini:2011kv}
\be
H_A=2\pi \int_{-1}^{1}dz \frac{1-z^2}{2} T_{tt}(z),
\ee  
where stress tensor is evaluated at $t=0$ slice. The stress tensor can be seperated into homomorphic and anti-holomorphic part. We will only consider chiral operators in this paper,  the holomorphic part of modular Hamiltonian is 
\be
H_A=-\int_{-1}^1dz \frac{1-z^2}{2}T(z), 
\ee 
where we used the convention $T(z)=-2\pi T_{tt}|_{\text{hol}}$, following \cite{cft}. Zero modes of modular Hamiltonian are operators in $D(A)$ which are annihilated by modular Hamiltonian \cite{Faulkner:2017vdd} 
\be
[H_A, Q_A]=0.
\ee
They are invariant under modular flow and have a gravitational dual which is evaluated at minimal surface in the bulk. Since modular Hamiltonian commutes with itself, modular Hamiltonian is also a zero mode. All zero modes form a closed algebra in $D(A)$.  Among all of the zero modes, OPE blocks are most intriguing. OPE blocks are the most natural objects in operator product expansion of two primary operators \footnote{The conformal weight of two operators $\mathcal{O}_1$ and $\mathcal{O}_2$ are assumed to be equal, $h_1=h_2$. When $h_1\not=h_2$, there is also a natural generalization of OPE blocks, see \cite{Das:2018ajg}. We will not discuss those objects in this work.}
\be
\mathcal{O}_1(z_1)\mathcal{O}_2(z_2)=\frac{1}{(z_1-z_2)^{h_1+h_2}}\sum_{\mathcal{O}}c_{\mathcal{O}} \ Q[\mathcal{O}].
\ee 
For interval $A$, it may have a closed form 
\be
Q_A[\mathcal{O}]=c_h \int_{-1}^1 (\frac{1-z^2}{2})^{h-1}\mathcal{O}(z),
\ee 
where $h$ is the conformal weight of primary operator $\mathcal{O}$. The coefficient $c_h$ is an unspecified  constant which depends on convention. OPE block is an eigenvector of Casimir operator $\mathcal{L}_2$ of $SL(2,R)$ with eigenvalue  $-h(h-1)$, therefore its correlation function with  other OPE blocks  in a disjoint region is an eigenfunction of Casimir operator $\mathcal{L}_2$ with the same eigenvalue $-h(h-1)$. Such kind of eigenfunction is conformal block \cite{Dolan:2000ut,Dolan:2003hv}
\be
G_h(\eta)=\eta^h \ {}_2F_1(h,h,2h,-\eta).
\ee In \cite{Long:2019fay,Long:2019pcv}, the author studied connected correlation functions
\be
\langle Q_A[\mathcal{O}_1]\cdots Q_A[\mathcal{O}_{m}]Q_B[\mathcal{O}_{m+1}]\cdots Q_B[\mathcal{O}_{m+n}]\rangle_c,\quad m,n\ge 1
\ee 
and concluded that these connected correlators are finite in general. When $n=1$, the correlator is conformal block 
\be
\langle Q_A[\mathcal{O}_1]\cdots Q_A[\mathcal{O}_{m}]Q_B[\mathcal{O}]\rangle_c=D[h_i,h]G_h(\eta)\label{conf}
\ee 
up to a constant $D[h_i,h]$.  $h_i$ is conformal weight of primary operator $\mathcal{O}_i$. We will justify \eqref{conf} in more details in Appendix \ref{AA}. The constant $D[h_i,h]$ encodes rich data of CFT.  It can be found by evaluating the corresponding integrals carefully for $m=1, 2$. The result is \cite{Long:2019pcv}
\bea
D[h_i,h_j]&=&c_{h_i}^2 \mathcal{N}_{h_i}\frac{2^{2-4h_i}\pi \Gamma(h_i)^2}{\Gamma(h_i+\frac{1}{2})^2}\delta_{ij},\label{dij}\\
D[h_1,h_2,h_3]&=&\frac{c_{h_1}c_{h_2}c_{h_3}C_{123}\pi^{3/2}(-1)^{h_3}\cos\frac{\pi}{2}(h_1+h_2-h_3) \Gamma(h_1)\Gamma(h_2)\Gamma(h_3)}{2^{h_1+h_2+h_3-3}\Gamma(\frac{1+h_1+h_2-h_3}{2})\Gamma(\frac{1+h_1+h_3-h_2}{2})\Gamma(\frac{1+h_2+h_3-h_1}{2})\Gamma(\frac{h_1+h_2+h_3}{2})}\label{d123}
\eea 
where $\mathcal{N}_{h_i}, C_{123}$ are constants in two and three point correlation functions
\bea
\langle \mathcal{O}_{i}(z_1)\mathcal{O}_j(z_2)\rangle&=&\frac{\mathcal{N}_{h_i}\delta_{ij}}{(z_1-z_2)^{2h_i}},\\
\langle\mathcal{O}_1(z_1)\mathcal{O}_2(z_2)\mathcal{O}_3(z_3)\rangle&=&\frac{C_{123}}{(z_1-z_2)^{h_1+h_2-h_3}(z_2-z_3)^{h_2+h_3-h_1}(z_1-z_3)^{h_1+h_3-h_2}}.
\eea 
The most general connected correlation function from OPE blocks is 
\bea
&&\langle Q_A[\mathcal{O}_1]\cdots Q_A[\mathcal{O}_{m_1}]Q_B[\mathcal{O}_{m_1+1}]\cdots Q_B[\mathcal{O}_{m_1+m_2}]Q_C[\mathcal{O}_{m_1+m_2+1}]\cdots Q_C[\mathcal{O}_{m_1+m_2+m_3}]\cdots\rangle_c,\nn\\
&&\hspace{180pt}m_1\ge m_2\ge\cdots \ge 0\label{young}
\eea 
where capital letters $A,B,C$ denote distinct spatial regions. We assume they are disjoint. A correlator \eqref{young} with $m_1$ OPE blocks in region A, $m_2$ OPE blocks in region B, etc. will be called $Y$-type, following the convention \cite{Long:2019pcv}.  We use a Young diagram 
\be
Y=(m_1,m_2,m_3,\cdots),\quad m_1\ge m_2\ge\cdots\ge1
\ee 
to label correlator \eqref{young}. Interestingly, the correlators of $(m,n)$-type with $m\ge n\ge1$ have been studied extensively in \cite{Long:2019fay, Long:2019pcv} while correlators of $(m)$-type have not been explored systematically though they correspond to one row Young diagrams, which should be much simpler than two row Young diagrams. This is partly because (m)-type correlators are divergent in general. However, we will show that they may always obey logarithmic law 
\be
\langle Q_A[\mathcal{O}_1]\cdots Q_A[\mathcal{O}_m]\rangle_c=C[h_i]\log\frac{L}{\epsilon},\label{corre} 
\ee 
where $\epsilon$ is a UV cutoff. 
Therefore the coefficients $C[h_i]$ are cutoff independent and encode rich information of the theory. The logarithmic behaviour of $(m)$-type correlator is similar to R\'enyi entropy \cite{Calabrese:2004eu}
\be
S_A^{(n)}=\frac{c}{12}(1+\frac{1}{n})\log\frac{L}{\epsilon}\label{renyi}
\ee 
for any CFT$_2$. Therefore, one should not discard \eqref{corre} just because they are divergent. Actually, R\'enyi entropy is a special combination of \eqref{corre} when all zero modes are modular Hamiltonians. To see this point, we define
\be
\tilde{S}_A^{(n)}\equiv (1-n)S_A^{(n)}=\log \text{tr}_A \rho_A^{n}=\log \text{tr}_A e^{-(n-1)H_A}\rho_A=\log\langle e^{-(n-1)H_A}\rangle.
\ee 
By expanding $\tilde{S}_A^{(n)}$ around $n=1$, we find \footnote{We insert a factor $(-1)^m$ in the expression, this changes the definition of connected correlation function in \cite{Long:2019fay} by a minus sign for odd $m$.}
\be
\tilde{S}_A^{(n)}=\sum_{m=1}^\infty \frac{(-1)^m}{m!}\langle H_A^m\rangle_c (n-1)^m.
\ee 
We conclude that R\'enyi entropy is a generator of connected correlation function of modular Hamiltonian, 
\be
\langle H_A^m\rangle_c=(-1)^m \frac{\partial^m \tilde{S}_A^{(n)}}{\partial n^m}|_{n=1}=\frac{c}{12}(\delta_{m,1}+m!)\log\frac{L}{\epsilon},\quad m\ge 1.\label{hm}
\ee 
$\langle H_A^m\rangle_c$ in \eqref{hm} is a special example of $(m)$-type correlators \eqref{corre}. We may regard \eqref{corre} as a generalization of logarithmic law of R\'enyi entropy. 

\section{Logarithmic behaviour of connected correlation function}\label{loglaw}
In this section, we will calculate $(m)$-type correlators in CFT$_2$ order by order. The lowest order is $m=1$. This is not fixed as one can always shift $Q_A[\mathcal{O}]$ by a constant, 
\be
Q'_A[\mathcal{O}]=Q_A[\mathcal{O}]+\text{const.}
\ee 
The constant term does not affect correlators with $m\ge2$ since its effects are canceled in connected correlation functions. However, it indeed affect $(1)$-type correlators. To determine $(1)$-type correlators, one should specify the constant term by other requirement. When $Q_A$ is modular Hamiltonian $H_A$, reduced density matrix  is normalized 
\be
\text{tr}_A\ \rho_A=1
\ee 
which may be used to fix the constant term.
However, for other OPE blocks, we don't find a nature condition to fix this constant term. One may choose this term to be 0. 
 In the following, we will consider the cases $m\ge 2$. 
\subsection{$(2)$-type}
$(2)$-type correlator is fixed by conformal symmetry 
\be
\langle Q_A[\mathcal{O}]Q_A[\mathcal{O}]\rangle_c=\frac{c_h^2\mathcal{N}_h}{2^{2h-2}}I_2[h]
\ee 
with 
\be
I_2[h]=\ :\int_{-1}^1dz_1\int_{-1}^1 dz_2 \frac{(1-z_1^2)^{h-1}(1-z_2^2)^{h-1}}{(z_1-z_2)^{2h}}:\label{i2h}.
\ee 
As \cite{Long:2019fay}, $:\cdots:$ means that one should regularize the integral since the integrand is divergent at $z_1=z_2$. We will omit $:\cdots:$ to simplify notation in the following. The method of regularization  has been justified by \cite{Long:2019fay,Long:2019pcv}, however, we'd like to evaluate it carefully since we will meet new features in the integral. The first step is to calculate $z_2$ integral as if there is no pole, then 
\be
I_2[h]=\frac{(-1)^{-h} \sqrt{\pi}\Gamma(h)}{\Gamma(h+\frac{1}{2})}\int_{-1}^1dz_1\frac{1}{1-z_1^2}.\label{I2h}
\ee 
The integral is still divergent once $z_1$ approaches the bondary of the interval, therefore, we insert a UV cutoff $\epsilon$ into the integral 
\be
I_2[h]=\frac{(-1)^{-h} \sqrt{\pi}\Gamma(h)}{\Gamma(h+\frac{1}{2})}\int_{-1+\epsilon}^{1-\epsilon}dz_1\frac{1}{1-z_1^2},\quad \epsilon\to0^+.\label{I2h2}
\ee 
The system is symmetric by exchanging the two end points, therefore the insertion of $\epsilon$ is chosen to be symmetric. Now \eqref{I2h2} is well defined, we find 
\be
I_2[h]=\frac{(-1)^{-h} \sqrt{\pi}\Gamma(h)}{\Gamma(h+\frac{1}{2})}\log\frac{2}{\epsilon}.\label{i2hh}
\ee 
Matching it with formula \eqref{corre} and using $L=2$, we find 
\be
C[h,h]=c_h^2\mathcal{N}_h\frac{(-1)^{-h} \sqrt{\pi}\Gamma(h)}{2^{2h-2}\Gamma(h+\frac{1}{2})}.\label{chh}
\ee 

\subsection{$(3)$-type}
(3)-type correlator is also fixed by conformal symmetry up to a structure constant of three point function, 
\be
\langle Q_A[\mathcal{O}_1]Q_A[\mathcal{O}_2]Q_A[\mathcal{O}_3]\rangle_c=\frac{c_{h_1}c_{h_2}c_{h_3}C_{123}}{2^{h_1+h_2+h_3-3}}I_3[h_1,h_2,h_3],
\ee 
with 
\be
I_3[h_1,h_2,h_3]=(\prod_{i=1}^3\int_{-1}^1 dz_i)\frac{(1-z_1^2)^{h_1-1}(1-z_2^2)^{h_2-1}(1-z_3^2)^{h_3-1}}{(z_1-z_2)^{h_1+h_2-h_3}(z_2-z_3)^{h_2+h_3-h_1}(z_1-z_3)^{h_1+h_3-h_2}}.\label{i3h123}
\ee
The integral can be evaluated for integer conformal weight case by case. We could check  that it always obeys logarithmic law
\be
I_3[h_1,h_2,h_3]=b_3[h_1,h_2,h_3]\log\frac{2}{\epsilon}
\ee 
up to $h_1,h_2,h_3\le 6$. The coefficients $b_3[h_1,h_2,h_3]$ are collected in Table \ref{tableof[3]type}. 
\begin{table}
	\centering
	\caption{$b_3[h_1,h_2,h_3]$}\begin{spacing}{1.19}
	\begin{tabular}{||c|c|c|c||c|c|c|c||c|c|c|c||c|c|c|c||}
		\hline
		$h_1$& $h_2$ & $h_3$& $b_3$&$h_1$& $h_2$ & $h_3$& $b_3$&$h_1$& $h_2$ & $h_3$& $b_3$&$h_1$& $h_2$ & $h_3$& $b_3$ \\
		\hline
		1&1&1&0&2&1&1&4&2&2&1&0&2&2&2&$-4$\\
		\hline
		3&1&1&0&3&2&1&$-\frac{8}{3}$&3&2&2&0&3&3&1&0\\
		\hline
		3&3&2&$\frac{32}{9}$ &    3&3&3&0&4&1&1&$\frac{8}{3}$&4&2&1&0\\
		\hline
		4&2&2&$\frac{16}{9}$&4&3&1&$\frac{32}{15}$&4&3&2&0&4&3&3&$-\frac{32}{9}$\\
		\hline
		4&4&1&0&4&4&2&$-\frac{16}{5}$&4&4&3&0&4&4&4&$\frac{64}{15}$
		\\\hline
		5&1&1&0&5&2&1&$-\frac{64}{45}$&5&2&2&0&5&3&1&0
		\\\hline
		5&3&2&$-\frac{64}{45}$&5&3&3&0&5&4&1&$-\frac{64}{35}$&5&4&2&0
		\\\hline
		5&4&3&$\frac{256}{75}$&5&4&4&0&5&5&1&0&5&5&2&$\frac{512}{175}$
		\\\hline
		5&5&3&0&5&5&4&$-\frac{1024}{225}$&5&5&5&0&6&1&1&$\frac{64}{15}$
		\\\hline
		6&2&1&0&6&2&2&$\frac{32}{45}$&6&3&1&$\frac{64}{63}$&6&3&2&0
		\\\hline
		6&3&3&$\frac{256}{225}$&6&4&1&0&6&4&2&$\frac{128}{105}$&6&4&3&0
		\\\hline
		6&4&4&$-\frac{256}{75}$&6&5&1&$\frac{512}{315}$&6&5&2&0&6&5&3&$-\frac{1024}{315}$
		\\\hline
		6&5&4&0&6&5&5&$\frac{8192}{1575}$&6&6&1&0&6&6&2&$-\frac{512}{189}$
		\\\hline
		6&6&3&0&6&6&4&$\frac{2048}{441}$&6&6&5&0&6&6&6&$-\frac{2048}{315}$
		\\\hline
	\end{tabular}\label{tableof[3]type}\end{spacing}
\end{table}
Note $b_3[h_1,h_2,h_3]$ is always 0 for $h_1+h_2+h_3$ odd. It is non-vanishing only  for $h_1+h_2+h_3$ even. However, it is hard to find a general formula from these results. The formula $C[h_1,h_2,h_3]$ is 
\be
C[h_1,h_2,h_3]=\frac{c_{h_1}c_{h_2}c_{h_3}C_{123}}{2^{h_1+h_2+h_3-3}}b_3[h_1,h_2,h_3]. \label{ch1h2h3}
\ee

\subsection{$(4)$-type}
Four point function of primary operators can only be fixed up to a function of cross ratio by global conformal invariance. However, it is much more simpler when four primary operators are identical and their conformal weight $h$ is an integer \cite{Datta:2014uxa},
\be
\langle\mathcal{O}(z_1)\mathcal{O}(z_2)\mathcal{O}(z_3)\mathcal{O}(z_4)\rangle=\frac{\mathcal{N}_h^2}{z_{12}^{2h}z_{34}^{2h}}f(\theta),
\ee 
where $z_{ij}=z_i-z_j$ and $\theta=x+\frac{1}{x}-2$, 
\be
\theta=\frac{z_{12}^2z_{34}^2}{z_{13}z_{14}z_{23}z_{24}},\quad x=\frac{z_{13}z_{24}}{z_{14}z_{23}}.
\ee 
The function $f(\theta)$ only depends on $\theta$ follows from the discrete symmetry 
 $z_1\leftrightarrow z_2$ or $z_3\leftrightarrow z_4$. It should be a polynomial of $\theta$ since there is no branch cut. The maximum degree of the polynomial is $2h$ since the most singular behaviour of the four point function is $z_{13}^{-2h}$ when $z_1\to z_3$. Therefore the function $f(\theta)$ is fixed up to a finite number of unkown constants \cite{Long:2014oxa}
  \be
 f(\theta)=\sum_{j=0}^{2h}a[h,j]\theta^j. \label{ftheta}
 \ee
 The first and last coefficient are fixed to 1 by examing the limit $z_1\to z_2$ and $z_1\to z_3$
 \be
 a[h,0]=a[h,2h]=1.
 \ee 
 Therefore we may consider the $(4)$-type correlator
 \be
 \langle Q_A[\mathcal{O}]^4\rangle_c=\frac{c_h^4\mathcal{N}_h^2}{2^{4h-4}}I_4[h,h,h,h], 
 \ee 
 where 
 \be
 I_4[h,h,h,h]=(\prod_{i=1}^4\int_{-1}^1dz_i (1-z_i^2)^{h-1})(\frac{f(\theta)}{z_{12}^{2h}z_{34}^{2h}}-\frac{1}{z_{12}^{2h}z_{34}^{2h}}-\frac{1}{z_{13}^{2h}z_{24}^{2h}}-\frac{1}{z_{14}^{2h}z_{23}^{2h}}).
 \ee 
 The last three terms are inserted because the correlator is connected. The basic integral is 
 \be
 {}_jI_4[h,h,h,h]=(\prod_{i=1}^4\int_{-1}^1dz_i (1-z_i^2)^{h-1})\frac{\theta^j}{z_{12}^{2h}z_{34}^{2h}},\quad j=0,1,\cdots,2h.
 \ee 
 The integrals from last three terms of $I_4[h,h,h,h]$ are $-3\times{}_0I_4[h,h,h,h]$. It turns out that the general structure of ${}_jI_4$ is 
 \be
 {}_jI_4[h,h,h,h]=b_4[h,j]\log\frac{2}{\epsilon}+b_4'[h,j]\log^2\frac{2}{\epsilon}.
 \ee  In Table \ref{[4]type2}, we calculate each basic integral ${}_jI_4$ up to $h\le 6$. 
 \begin{table}
 	\caption{$b_4[h,j]$ and $b_4'[h,j]$}
 	\begin{spacing}{1.39}	\begin{tabular}{||c|c|c|c||c|c|c|c||}
 			\hline
 			h&j&$b_4[h,j]$&$b_4'[h,j]$&	h&j&$b_4[h,j]$&$b_4'[h,j]$\\\hline
 			1&0&0&4&1&1&$\frac{4\pi^2}{3}$&0\\\hline1&2&$-\frac{8\pi^2}{3}$&8&2&0&0&$\frac{16}{9}$
 			\\\hline	2&1&$\frac{16(15+2\pi^2)}{45}$&0&2&2&$\frac{32(15-\pi^2)}{45}$&0\\\hline2&3&$\frac{16(-75+8\pi^2)}{45}$&0&2&4&$\frac{64(60-7\pi^2)}{45}$&$\frac{32}{9}$\\\hline
 			3&0&0&$\frac{256}{225}$&3&1&$\frac{64(245+24\pi^2)}{2835}$&0\\\hline3&2&$\frac{128}{9}$&0&3&3&$\frac{128(-35+4\pi^2)}{315}$&0
 			\\\hline
 			3&4&$\frac{256(119-12\pi^2)}{567}$&0&3&5&$\frac{64(-315+32\pi^2)}{105}$&0\\\hline3&6&$\frac{128(1715-176\pi^2)}{315}$&$\frac{512}{225}$&4&0&0&$\frac{1024}{1225}$\\\hline4&1&$\frac{512(7007+600\pi^2)}{675675}$&0&4&2&$\frac{512(1001+24\pi^2)}{27027}$&0\\\hline4&3&$\frac{8192\pi^2}{5005}$&0&4&4&$\frac{1024(5005-492\pi^2)}{135135}$&0\\\hline4&5&$\frac{512(-35035+3552\pi^2)}{135135}$&0& 4&6&$\frac{512(71071-7200\pi^2)}{75075}$&0\\\hline4&7&$\frac{1024(-1184183+120000\pi^2)}{675675}$&0&4&8&$\frac{8192(553553-56145\pi^2)}{675675}$&$\frac{2048}{1225}$\\\hline
 			5&0&0&$\frac{65536}{99225}$&5&1&$\frac{4096(1485341+117600\pi^2)}{1206079875}$&0\\\hline5&2&$\frac{8192(70499+2400\pi^2)}{24613875}$&0&
 			5&3&$\frac{8192(17017+1920\pi^2)}{6891885}$&0\\\hline5&4&$\frac{16384(17017-960\pi^2)}{6891885}$&0&5&5&$\frac{8192(-85085+8688\pi^2)}{6891885}$&0\\\hline5&6&$\frac{16384(556699-56400\pi^2)}{24613875}$&0&5&7&$\frac{16384(-14413399+1460400\pi^2)}{172297125}$&0\\\hline5&8&$\frac{32768(2901679-294000\pi^2)}{18555075}$&0&5&9&$\frac{4096 (-1137443021 + 115248000 \pi^2)}{241215975}$&0
 			\\\hline
 			5&10&$\frac{8192 (431441725-43717632 \pi^2)}{48243195}$&$\frac{131072}{99225}$&6&0&0&$\frac{262144}{480249}$\\\hline6&1&$\frac{32768(2817529+211680\pi^2)}{19249034805}$&0&6&2&$\frac{32768(600457+23520\pi^2)}{712927215}$&0\\\hline6&3&$\frac{131072(46189+3360\pi^2)}{130945815}$&0&6&4&$\frac{65536}{945}$&0
 			\\\hline
 			6&5&$\frac{65536(-46189+6000\pi^2)}{43648605}$&0&6&6&$\frac{65536(54587-5520\pi^2)}{11904165}$&0\\\hline6&7&$\frac{1048576(-3553+360\pi^2)}{3357585}$&0&6&8&$\frac{131072 (1230307-124656 \pi^2)}{38886939}$&0	\\\hline
 			6&9&$\frac{32768(-3059513171+309993600\pi^2)}{6416344935}$&0&6&10&$\frac{32768 (292487767-29635200 \pi^2)}{161756595}$&0\\\hline6&11&$\frac{131072 (-45126653+4572288 \pi^2)}{26189163}$&0&6&12&$\frac{65536(12092141633 -1225197120\pi^2 )}{916620705}$&$\frac{524288}{480249}$
 			\\\hline
 	\end{tabular}\end{spacing}\label{[4]type2}
 \end{table}
 
 Several remarks: 
 \begin{enumerate}
 	\item $b_4'[h,j]$ is non-zero only for $j=0$ and $2h$, they are proportional to each other 
 	\be
 	b_4'[h,2h]=2b_4'[h,0].
 	\ee 
 	Therefore their contribution to connected correlation function is canceled in $I_4[h,h,h,h]$. 
 	\item One can read $C[h,h,h,h]$ in \eqref{corre}
 	\be
 	C[h,h,h,h]=\frac{c_h^4\mathcal{N}_h^2}{2^{4h-4}}\sum_{j=1}^{2h}a[h,j]b_4[h,j].\label{chhhh}
 	\ee  We will return to this formula in the following sections. 
 	
 \end{enumerate}
 
 \section{Massless free scalar theory}\label{freescalar}
 In Section \ref{loglaw}, we test the formula \eqref{corre} for $2\le m\le 4$ in general CFT$_2$. In this section, we will extend these results to higher orders for massless free scalar theory. In this theory, modular Hamiltonian can be evaluated in momentum space \cite{Long:2019fay}
\be
H_A=2\pi \sum_v v b_v^\dagger b_v+\text{const}, 
\ee 
where $b_v,b_v^\dagger$ are annihilation and creation operators by quantizing the theory in subregion $D(A)$. They obey standard commutation relations. The constant term is fixed by normalization of reduced density matrix $\rho_A=e^{-H_A}$
\be
\text{tr}_A \ \rho_A=1.\ee
R\'enyi entropy are found from momentum space \footnote{We just consider left-moving part.}, 
\be
S^{(n)}_A=\frac{\pi}{12}(1+\frac{1}{n})\delta(0).\label{renyifree}
\ee 
The Dirac delta function $\delta(0)$ is divergent in momentum space 
\be
\delta(0)=\delta(v-v')|_{v'=v}.
\ee 
The author \cite{Long:2019fay} noticed that it matches with \eqref{renyi} if one regularizes the Dirac delta function by a UV cutoff $\epsilon$,
\be
\delta(0)\to \frac{1}{\pi}\log\frac{2}{\epsilon}\label{div}
\ee 
in real position space.
For massless free scalar theory, there are infinite many primary conserved currents \cite{Bakas:1990ry}
\be
J_s=\sum_{k=1}^{s-1}A_k^s \partial^k\phi\partial^{s-k}\phi^*,\quad A_k^s=(-1)^k C_{s-1}^k C_{s-1}^{s-k}.
\ee 
We will study real scalar, therefore $s$ is even, 
\be
s=2,4,6,\cdots.
\ee 
The conformal weight of $J_s$ is equal to $s$, therefore the corresponding OPE block is\be
Q_A[J_s]=c_s\int_{-1}^1 (\frac{1-z^2}{2})^{s-1}J_s(z).\label{QAJs}
\ee 
They have rather simple form in momentum space \cite{Long:2019pcv} 
\be
Q_A[J_s]=2\pi \sum_v f_s(v)b_v^\dagger b_v+\text{const.}\label{qajs}
\ee 
where 
\be
f_s(v)=\frac{c_s}{2\pi v}S(s;iv,-iv).
\ee 
The function $S(s;a,b)$ is a generalized hypergeometric function 
\be
S(s;a,b)=-a(s-1)\frac{\Gamma(1+b)}{\Gamma(2+b-s)}{}_3F_2(1-a,1-s,2-s;2,2+b-s;1).
\ee 
For integer $s$, it is a polynomial of two variables $a$ and $b$. The normalization constant $c_s$ is chosen such that 
\be
f_s(v)\to v^{s-1},\quad v\to \infty. 
\ee 
Since OPE blocks are zero modes of modular Hamiltonian, one can define deformed reduced density matrix \cite{Long:2019pcv}
\be
\rho_A(\alpha_i)=e^{-\sum_{s\ge 2,\text{even}}\alpha_s Q_A[J_s]}=e^{-2\pi \sum_{s\ge 2}\alpha_s f_s(v)b_v^\dagger b_v},
\ee 
which is related to the generator of connected correlation function of OPE blocks by
\be
T_A(\alpha_i)=\log \langle \rho_A(\alpha_i)\rangle.
\ee 
The deformed reduced density matrix is not normalized in this work. The generator is similar to partition function of a thermodynamic systerm. Expanding $T_A(\alpha_i)$ for small $\alpha_i$, we could find its relation to connected correlation function
\be
T_A(\alpha_i)=\sum_{m_1,m_2,\cdots}\langle Q_A[J_{s_1}]^{m_1}Q_A[J_{s_2}]^{m_2}\cdots \rangle_c \prod_{i=1}\frac{(-\alpha_i)^{m_i}}{m_i!}.
\ee 
For the $Q_A[J_s]$ in \eqref{QAJs}, $T_A(\alpha_i)$ has an exact form 
\bea
T_A(\alpha_i)&=&\log \langle \rho_A(\alpha_i)\rangle=\log \text{tr}_A e^{-H_A-\sum_{s\ge 2}\alpha_s Q_A[J_s]}=\log \text{tr}_A e^{-2\pi \sum_v h(\alpha_i;v)b_v^\dagger b_v},\nn\\&=&-\int_0^\infty dv \log\frac{1-e^{-2\pi h(\alpha_i;v)}}{1-e^{-2\pi v}}\delta(0),
\eea 
where 
\be
h(\alpha_i;v)=v+\sum_{s} \alpha_s f_s(v).
\ee 
Dirac delta function has the same origin in momentum space  as \eqref{renyifree}. Therefore using the rule \eqref{div}, 
\be
T_A(\alpha_i)=F(\alpha_i)\log\frac{2}{\epsilon},\label{TAa}
\ee 
where 
\be
F(\alpha_i)=-\frac{1}{\pi}\int_0^\infty dv \log\frac{1-e^{-2\pi h(\alpha_i;v)}}{1-e^{-2\pi v}}.\label{Fai}
\ee 
\eqref{TAa} should be regarded as a ``R\'enyi entropy'' corresponds to deformed reduced density matrix. It still obeys logarithmic law. For $\alpha_2=n-1,\alpha_i=0,i>2$, it reproduces formula \eqref{renyi} for $c=1$,
\be
\frac{T_A(\alpha_i)}{1-n}|_{\alpha_2=n-1,\alpha_{i>2}=0}=S^{(n)}_A.
\ee 
Several interesting properties appear in \eqref{TAa}.
\begin{enumerate}
	\item The formula \eqref{TAa} is equivalent to \eqref{corre} for free scalar theory. Note the primary operator is quadratic in terms of scalar field in this example. The connected correlation function coefficient $C$ is
	\be
	C[s_i,m_i]=(\prod_{j}(-1)^{m_j}\frac{\partial^{m_j}}{\partial \alpha_j^{m_j}})F(\alpha_i)|_{\alpha_i=0}.
	\ee 
	Assume $\sum_i m_i=m\ge1$, then
	\be
	C[s_i,m_i]=2(-1)^{m-1}\int_0^\infty dv   \prod_{j}f_{s_j}^{m_j}(v) \times \frac{\partial^{m-1}}{\partial v^{m-1}}\frac{1}{e^{2\pi v}-1}.
	\ee 
	The function $f_s(v)$ and the particle number $N(v)=\frac{1}{e^{2\pi v}-1}$ have the following asymptotic behaviour 
	\be
	f_s(v)\sim \left\{
	\begin{aligned}
	\text{const.}\ v,\quad v\to 0, \\
		\ v^{s-1},\quad v\to\infty, 
	\end{aligned}
	\right.
	\quad N(v)\sim \left\{
	\begin{aligned}
		\frac{1}{2\pi v},\quad v\to 0, \\
		\ e^{-2\pi v},\quad v\to\infty. 
	\end{aligned}
	\right.
	\ee 
	Therefore one can integrate by parts, 
	\be
	C[s_i,m_i]=2\int_0^\infty dv\ \frac{1}{e^{2\pi v}-1} \times \frac{\partial^{m-1}}{\partial v^{m-1}} \prod_{j} f_{s_j}^{m_j}(v).
	\ee 
	The coefficients up to $m\le 4$ and spin less than 6 are collected in Table \ref{ccoe}.
	 \begin{table}
		\caption{$C$ function for free scalar}
		\begin{spacing}{1.29}	\begin{tabular}{||c|c|c|c||c|c|c|c||c|c|c|c||}
				\hline
				$m_1$&$m_2$&$m_3$&$C[m_1,m_2,m_3]$&$m_1$&$m_2$&$m_3$&$C[m_1,m_2,m_3]$&$m_1$&$m_2$&$m_3$&$C[m_1,m_2,m_3]$\\\hline
				1&0&0&$\frac{1}{12}$&2&0&0&$\frac{1}{6}$&3&0&0&$\frac{1}{2}$\\\hline
				4&0&0&2&0&1&0&$-\frac{1}{120}$&1&1&0&0\\\hline2&1&0&$\frac{1}{15}$&3&1&0&$\frac{3}{5}$&0&2&0&$\frac{3}{175}$
				\\\hline
				1&2&0&$\frac{3}{25}$&2&2&0&$\frac{76}{75}$&0&3&0&$\frac{27}{125}$\\\hline 1&3&0&$\frac{288}{125}$&0&4&0&$\frac{4932}{625}$&0&0&1&$\frac{1}{168}$\\\hline 1&0&1&0&2&0&1&$-\frac{1}{63}$&3&0&1&$\frac{1}{21}$\\\hline 0&1&1&0&1&1&1&$\frac{2}{35}$&2&1&1&$\frac{302}{315}$\\\hline 0&2&1&$\frac{12}{35}$&1&2&1&$\frac{852}{175}$&0&3&1&$\frac{23568}{875}$\\\hline 0&0&2&$\frac{40}{1617}$&1&0&2&$\frac{40}{147}$&2&0&2&$\frac{4828}{1323}$
				\\\hline 0&1&2&$\frac{64}{49}$&1&1&2&$\frac{5064}{245}$&0&2&2&$\frac{181008}{1225}$\\\hline 0&0&3&$\frac{2320}{343}$&1&0&3&$\frac{124720}{1029}$&0&1&3&$\frac{1127456}{1029}$\\\hline0&0&4&$\frac{224805760}{21609}$&&&&&&&&
				\\\hline
				\end{tabular}\end{spacing}\label{ccoe}\end{table}

In the table, $m_1,m_2,m_3$ are number of spin 2,4,6, respectively. In the following we will match them with corresonding results of Section \ref{loglaw}. 
\begin{enumerate}
	\item $(2)$-type.  we need normalization constants 
\bea
&&N_2=\frac{1}{8\pi^2},\quad N_4=\frac{135}{2\pi^2},\quad N_6=\frac{378000}{\pi^2},\\
&&c_2=-2\pi,\quad c_4=\frac{2\pi}{15},\quad c_6=-\frac{\pi}{105}.
\eea 
We reproduce $(2)$-type correlators of Table \ref{ccoe} using formula \eqref{chh}. 
\item $(3)$-type.
Three point function coefficients are found in Table \ref{three}. Using these coefficients, formula \eqref{ch1h2h3} and Table \ref{tableof[3]type}, we reproduce $(3)$-type correlators of Table \ref{ccoe}. 
\begin{table}
\centering	\caption{Three point function for free scalar}
	\begin{spacing}{1.29}
		\begin{tabular}{||c|c|c|c||c|c|c|c||c|c|c|c||c|c|c|c||}\hline
			$s_1$&$s_2$&$s_3$&$C_{s_1,s_2,s_3}$&	$s_1$&$s_2$&$s_3$&$C_{s_1,s_2,s_3}$&	$s_1$&$s_2$&$s_3$&$C_{s_1,s_2,s_3}$&	$s_1$&$s_2$&$s_3$&$C_{s_1,s_2,s_3}$
				\\\hline
				2&2&2&$\frac{1}{8\pi^3}$&2&2&4&$\frac{9}{4\pi^3}$&2&2&6&$\frac{75}{\pi^3}$&
				2&4&4&$\frac{135}{\pi^3}$\\\hline2&4&6&$\frac{9450}{\pi^3}$&2&6&6&$\frac{1134000}{\pi^3}$&4&4&4&$\frac{10935}{\pi^3}$& 4&4&6&$\frac{1215000}{\pi^3}$\\\hline4&6&6&$\frac{190512000}{\pi^3}$&6&6&6&$\frac{39463200000}{\pi^3}$&&&&&&&&
				\\\hline
			\end{tabular}\end{spacing}\label{three}\end{table}
\item The coefficients $a[h,j]$ in function \eqref{ftheta} are given in Table \ref{ahj}.
 \begin{table}
\centering	\caption{$a[h,j]$ for free scalar}
	\begin{spacing}{1.29}\begin{tabular}{||c|c|c||c|c|c||c|c|c||c|c|c||c|c|c||}
			\hline h&j&$a[h,j]$&h&j&$a[h,j]$&h&j&$a[h,j]$&	h&j&$a[h,j]$&	h&j&$a[h,j]$\\\hline
		2&0&1&2&1&8&2&2&10&2&3&4&2&4&1\\\hline
		4&0&1&4&1&32&4&2&$\frac{1928}{5}$&4&3&$\frac{3856}{5}$&4&4&$\frac{2738}{5}$\\\hline 4&5&176&4&6&52&4&7&8&4&8&1& 6&0&1\\\hline6&1&72&6&2&3756&6&3&$\frac{587980}{21}$&6&4&$\frac{513210}{7}$&6&5&$\frac{640564}{7}$\\\hline6&6&$\frac{1311514}{21}$& 6&7&24732&6&8&5805&6&9&760&6&10&126\\\hline 6&11&12&6&12&1&&&&&&&&&
		\\\hline
	
\end{tabular}\end{spacing}\label{ahj}\end{table}
Using formula \eqref{chhhh} and Table \ref{[4]type2}, we obtain 
\be
C[2,2,2,2]=2,\quad C[4,4,4,4]=\frac{4932}{625},\quad C[6,6,6,6]=\frac{224805760}{21609}.
\ee 
They are exactly the corresponding coefficients of Table \ref{ccoe}. 
\end{enumerate}
\item The parameters are constrained in the region $\alpha_i\ge0$ since the exponential operator should be bounded for large $v$. Interestingly, for $ i>2$, $\alpha_i$ also have upper bound. For example, we consider $\alpha_4\not=0$, in this case 
\be
h(\alpha_2,\alpha_4;v)=(\alpha_2+1)v+\alpha_4(v^3-\frac{1}{5}v).
\ee 
It should be non-negative for all $v$, otherwise the integrand for $F(\alpha_i)$ is not real. 
\be
h(\alpha_2,\alpha_4;v)\ge0,\quad \forall v>0.
\ee 
Therefore we find 
\be
\alpha_4\le 5(1+\alpha_2).
\ee 
\item Connected correlation functions are related to small $\alpha_i$ expansion of generator $T_A(\alpha_i)$. One may also regard $T_A(\alpha_i)$ as an independent quantity. As has been pointed out in \cite{Long:2019pcv}, one can keep $\alpha_s (s\not=2)$ finite while 
\be
\alpha_2\to\infty,
\ee 
then 
$T_A(\alpha_i)$ is universal
\be
F(\alpha_i)=-\frac{1}{12},
\ee 
which is independent of $\alpha_s$. 
\end{enumerate}

\section{Analytic continuation of Selberg integral}
In Section \ref{freescalar}, we use the example of free scalar to justify the logarithmic behaviour derived from conformal field theory in Section \ref{loglaw}. In this section, we use a rather different approach, analytic continuation of  Selberg integral  \cite{Selberg:1944}, to validate our method of  regularization. Selberg integral is a direct multi-dimensional generalization of integral representation of Beta function. The basic Selberg integral is 
\bea
S_n(\alpha,\beta,\gamma)&=&\prod_{i=1}^n \int_0^\infty dt_i \prod_{i=1}^n t_i^{\alpha-1}(1-t_i)^{\beta-1}\prod_{1\le i<j\le n}|t_i-t_j|^{2\gamma}\nn\\
&=&\prod_{j=0}^{n-1}\frac{\Gamma(\alpha+j \gamma)\Gamma(\beta+j\gamma)\Gamma(1+(j+1)\gamma)}{\Gamma(\alpha+\beta+(n+j-1)\gamma)\Gamma(1+\gamma)}.\label{SN}
\eea 
It is convergent in the region 
\be
\text{Re}(\alpha)>0,\text{Re}(\beta)>0,\text{Re}(\gamma)>-\text{min}(\frac{1}{n},\frac{\text{Re}(\alpha)}{n-1},\frac{\text{Re}(\beta)}{n-1}).
\ee 
The development of Selberg integral can be found in \cite{For}. 
\begin{enumerate}
	\item $(2)$-type. The integral \eqref{i2h} is  a ``Selberg integral'' for integer conformal weight, 
\be
n=2, \alpha=h,\beta=h,\gamma=-h.
\ee 
However, $\gamma$ is in the wrong region since \eqref{i2h} is divergent. Nevertheless, this provides a new way to regularize $I_2[h,h]$. By transforming variables  $z_i $ to $t_i$,
\be
z_i=2t_i-1
\ee 
we find 
\be
I_2[h,h]=4^{h-1}S_2(h,h,-h+\tilde{\epsilon}_2).
\ee 
We insert a $\tilde{\epsilon}_2$ since the integral is divergent, we  continue Selberg integral to $\tilde{\epsilon}_2\to0$
\be
I_2[h,h]=\frac{\Gamma(\frac{1}{2}-h)\Gamma(h)}{2\sqrt{\pi}\tilde{\epsilon_2}}+\cdots
\ee 
This reproduces \eqref{i2hh} by identifying 
\be
\frac{1}{\tilde{\epsilon}_2}=2 \log\frac{2}{\epsilon}.
\ee 
\item $(3)$-type. The general integral $I_3[h_1,h_2,h_3]$ \eqref{i3h123} is not the basic Selberg integral \eqref{SN}. However, when $h_1=h_2=h_3=h$ and even, 
\be
I_3[h,h,h]=8^{h-1}S_3(h,h,-\frac{h}{2}+\tilde{\epsilon}_3)=\frac{2^{3h-4}(-1)^{h/2}\Gamma(\frac{h}{2})^3}{3\Gamma(\frac{3h}{2})}\times \frac{1}{\tilde{\epsilon_3}}+\cdots.
\ee 
If we identify 
\be
\frac{1}{\tilde{\epsilon}_3}=6\log\frac{2}{\epsilon},
\ee 
then 
\be
I_3[h,h,h]=\frac{2^{3h-3}(-1)^{h/2}\Gamma(\frac{h}{2})^3}{\Gamma(\frac{3h}{2})}\log\frac{2}{\epsilon}.\label{i3hhh}
\ee 
This formula leads to 
\be
b_3[2,2,2]=-4,\quad b_3[4,4,4]=\frac{64}{15},\quad b_3[6,6,6]=-\frac{2048}{315},
\ee 
which could also be found from Table \ref{tableof[3]type}.
\item $(4)$-type. One can use Selberg integral only for $j=\frac{2h}{3}$, the conformal weight $h=3 k, k\in \mathcal{Z}^+$, then the following basic integral 
\be
{}_{2k}I_4[3k,3k,3k,3k]=2^{4(3k-1)}S_{4}(3k,3k,-k+\tilde{\epsilon}_4)=\frac{8^{4k-2}\Gamma(k)^4}{9\Gamma(4k)}\frac{1}{\tilde{\epsilon_4}}+\cdots.
\ee 
By identifying 
\be
\frac{1}{\tilde{\epsilon_4}}=12\log\frac{2}{\epsilon}, 
\ee 
we find 
\be
b_4[3,2]=\frac{128}{9},\quad b_4[6,4]=\frac{65536}{945}.
\ee 
A slightly more general $b_4$ is
\be
b_4[3k,2k]=\frac{4^{6k-2}\Gamma(k)^4}{3\Gamma(4k)}.
\ee 
\end{enumerate}
We will comment on analytic continuation of Selberg integral. The advantage of analytic continuation of Selberg integral is obvious. However, the disadvantages are also serious . Firstly, one cannot obtain all $(m)$-type correlators by using Selberg integral, even for $m=3$. A basic Selberg integral has only three parameters $a,b,c$ for a fixed $n$. To use Selberg integral, the conformal weight should be equal. The degree of $z_i-z_j$ should be the same. This restricts the applications of Selberg integral.  One should develop Selberg integral further. A general correlation function is  invariant under spacetime translation, then it will only depends on the distance of positions. Therefore, we propose a more general integral 
\be
S_n(\vec{\alpha},\vec{\beta};\bm{\gamma})=\prod_{i=1}^n \int_0^1 dt_i \prod_{j=1}^n t_j^{\alpha_j-1}(1-t_j)^{\beta_j-1} \prod_{1\le k<\ell\le n}|t_k-t_{\ell}|^{\gamma_{k\ell}},\label{SNG}
\ee 
where $\vec{\alpha},\vec{\beta}$ are vectors whose elements are 
\be
\vec{\alpha}_i=\alpha_i,\quad \vec{\beta}_i=\beta_i.
\ee 
The matrix $\bm{\gamma}$ is a $n\times n$ symmetric matrix whose diagonal terms are zero, 
\be
\bm{\gamma}_{ij}=\left\{\begin{aligned}\gamma_{ij},\quad i\not=j\\0,\quad i=j.\end{aligned}\right.
\ee 
The total number of parameters of $S_n(\vec{\alpha},\vec{\beta},\bm{\gamma})$ is 
\be
2n+\frac{n(n-1)}{2}=\frac{n(n+3)}{2}.
\ee 
Secondly, we should introduce a small paramter $\tilde{\epsilon}_i$ to control the divergent behaviour for analytic continuation of Selberg integral, the relationship between $\tilde{\epsilon}_i$ and UV cutoff $\epsilon$ is unclear without matching several examples.

\section{Analytic continuation of conformal block}\label{analycb}
Motivated by the success of analytic continuation of Selberg integral, we develop another method to overcome several difficulties in previous sections. The method is similar to point-splitting regularization in quantum field theory. For example, to extract the singular and finite part the expectation value of a quadratic operator $(\partial\phi)^2$, one should split the operator to $\partial\phi(z+\epsilon)\partial\phi(z)$ firstly, and then take the limit $\epsilon\to 0$, 
\be
\langle (\partial\phi)^2(z)\rangle=\lim_{\epsilon\to0}\langle\partial\phi(z+\epsilon)\partial\phi(z)\rangle.
\ee 
The $(m)$-type connected correlation function of OPE block is similar, we first split the region $A$ to two disjoint regions $A$ and $B$, 
\be
\langle Q_A[\mathcal{O}_1]\cdots Q_A[\mathcal{O}_m]\rangle_c\xrightarrow{split}\langle Q_A[\mathcal{O}_1]\cdots Q_A[\mathcal{O}_{m-1}]Q_B[\mathcal{O}_m]\rangle_c.\label{split}
\ee 
The right hand side of \eqref{split} is a $(m-1,1)$-type correlator, which has been shown to be finite \cite{Long:2019pcv} and proportional to conformal block, see equation \eqref{conf}. Formally,  we take the limit $B\to A$, then $(m-1,1)$-type correlator should be $(m)$-type correlator, 
\bea
\langle Q_A[\mathcal{O}_1]\cdots Q_A[\mathcal{O}_m]\rangle_c&=&\lim_{B\to A}\langle Q_A[\mathcal{O}_1]\cdots Q_A[\mathcal{O}_{m-1}]Q_B[\mathcal{O}_m]\rangle_c\nn\\
&=&\lim_{B\to A} D[h_1,\cdots,h_m]G_{h_m}(\eta).\label{split2}
\eea 
The right hand side of \eqref{split2} is only a function of cross ratio 
\be
\eta=\frac{x_{12}x_{34}}{x_{14}x_{23}},
\ee 
it approaches $-1$ when $B\to A$. The way to approach $A$ is subtle, we choose a way which is symmetric \footnote{There is another way to take the limit to get the same logarithmic behaviour. We will discuss it in Appendix \ref{analy}.}. More explicitly, the end points of $A$ and $B$ are parameterized as 
\be
x_1=1,\quad x_2=-1,\quad, x_3=1-\epsilon,\quad x_4=-1+\epsilon, \quad \epsilon>0.
\ee 
The parameter $\epsilon$ characterizes the distance between point $x_3$ to $x_1$, therefore it should be the same UV cutoff in Section \ref{loglaw}. 
The cross ratio is
\be
\eta=\frac{4(-1+\epsilon)}{(2-\epsilon)^2}\approx\ -1+\frac{\epsilon^2}{4}.
\ee 
The conformal block $G_h(\eta)$ presents logarithmic behaviour
\be
G_h(\eta)=\eta^h \ {}_2F_1(h,h,2h,-\eta)= \frac{2(-1)^h\Gamma(2h)}{\Gamma(h)^2}\log\frac{2}{\epsilon}+\cdots.\label{cbex}
\ee 
The terms in $\cdots$ are finite and suppressed in the limit $\epsilon\to0$, therefore 
\be
\langle Q_A[\mathcal{O}_1]\cdots Q_A[\mathcal{O}_m]\rangle_c=\frac{2(-1)^{h_m}\Gamma(2h_m)}{\Gamma(h_m)^2}D[h_i]\log\frac{2}{\epsilon}.
\ee 
It is intriguing that logarithmic behaviour in Section \ref{loglaw} emerges as analytic continuation of conformal block while the latter is fixed by conformal symmetry. We read a neat relation between $C[h_i]$ and $D[h_i]$ 
\be
C[h_i]=\frac{2(-1)^{h_m}\Gamma(2h_m)}{\Gamma(h_m)^2}D[h_i].\label{cdrel}
\ee 
In the following we will discuss the consequences of \eqref{cdrel}. 
\begin{enumerate}
	\item $(2)$-type. $D[h_i,h_j]$ can be found in \eqref{dij}, then 
	\be
	C[h,h]=\frac{c_h^2\mathcal{N}_h(-1)^h  2^{3-4h}\pi\Gamma(2h)}{\Gamma(h+\frac{1}{2})^2}.
	\ee 
	It is indeed \eqref{chh} for integer conformal weight. 
	\item $(3)$-type. $D[h_1,h_2,h_3]$ can be read from \eqref{d123}, then 
	\be
 C[h_1,h_2,h_3]=\frac{c_{h_1}c_{h_2}c_{h_3}C_{123}\pi^{3/2}(-1)^{\frac{h_1+h_2+h_3}{2}}\Gamma(h_1)\Gamma(h_2)\Gamma(h_3)\kappa}{2^{h_1+h_2+h_3-3}\Gamma(\frac{1+h_1+h_2-h_3}{2})\Gamma(\frac{1+h_1+h_3-h_2}{2})\Gamma(\frac{1+h_2+h_3-h_1}{2})\Gamma(\frac{h_1+h_2+h_3}{2})}.\label{cch1h2h3}
	\ee 
	where 
	\be
	\kappa=\frac{1}{2}[1+(-1)^{h_1+h_2+h_3}].
	\ee 
	for any integer $h_1,h_2,h_3$. We have written $C[h_1,h_2,h_3]$ symmetrically under exchange of any two coonformal weights. Interestingly, from \eqref{cch1h2h3} we find 
	\be
	b_3[h_1,h_2,h_3]=\frac{\pi^{3/2}(-1)^{\frac{h_1+h_2+h_3}{2}}\Gamma(h_1)\Gamma(h_2)\Gamma(h_3)\kappa}{\Gamma(\frac{1+h_1+h_2-h_3}{2})\Gamma(\frac{1+h_1+h_3-h_2}{2})\Gamma(\frac{1+h_2+h_3-h_1}{2})\Gamma(\frac{h_1+h_2+h_3}{2})}
	\ee  which should be very hard to evaluate in Section \ref{loglaw}.  It reproduces results of Table \ref{tableof[3]type}. When $h_1=h_2=h_3=h$ and even, 
	\be
	b_3[h,h,h]=\frac{(-1)^{\frac{h}{2}}\pi^{3/2}\Gamma(h)^3}{\Gamma(\frac{1+h}{2})^3\Gamma(\frac{3h}{2})}.
	\ee It is the same as \eqref{i3hhh} which is obtained from analytic continuation of Selberg integral.
	\item $(4)$-type. We can read $D[2,2,2,2]$ from \cite{Long:2019fay}
	\be
	D[2,2,2,2]=\frac{c}{6}\Rightarrow C[2,2,2,2]=2c.\label{x2222}
	\ee 
	This is valid for general CFT$_2$. This is consistent with \eqref{chhhh}. It is easy to find
	\bea
&&	N_2=\frac{c}{2},\quad c_2=-1,\nn\\
&&a[2,0]=1,\quad 	a[2,1]=\frac{8}{c},\quad a[2,2]=2+\frac{8}{c},\quad a[2,3]=4,\quad a[2,4]=1.
	\eea 
	Substituting them into \eqref{chhhh}, 
	\be
	C[2,2,2,2]=2c
	\ee 
	which is is the same as \eqref{x2222}. 
	\item $(m)$-type. For $m>4$, we can also read \cite{Long:2019fay}
	\be
	D[2^m]=\frac{m!c}{144}\Rightarrow C[2^m]=\frac{m! c}{12}.
	\ee 
	This is the same as \eqref{hm} for $m\ge 2$, which is obtained from R\'enyi entropy.
	\item $(1)$-type is very special, in this case, there is only one OPE block, there is no way to use the method of analytic continuation. 
	\item When all the OPE blocks commute with each other, then $C[h_i]$ should be independent of the order of conformal weight. There are $m$ ways to uplift $(m)$-type correlator to $(m-1,1)$-type,  the final result should be the same 
	\be
	C[h_1,\cdots,h_m]=\frac{2(-1)^{h_i}\Gamma(2h_i)}{\Gamma(h_i)^2}D[h_1,\cdots,h_{i-1},h_{i+1},\cdots,h_m,h_i], \quad i=1,\cdots,m.
	\ee This identity can be checked for free scalar \cite{Long:2019pcv}. For example, 
	\bea
	&&D[2,2,2,4]=\frac{3}{1400},\quad D[2,2,4,2]=\frac{1}{20}.
	\eea 
	Both of them lead to 
	\be
	C[2,2,2,4]=C[2,2,4,2]=\frac{2(-1)^4\Gamma(8)}{\Gamma(4)^2}D[2,2,2,4]=\frac{2(-1)^2\Gamma(4)}{\Gamma(2)^2}D[2,2,4,2]=\frac{3}{5}.
	\ee The reader can also check other examples in \cite{Long:2019pcv}. 
\item Coefficient $D[h_i]$ characterizes the leading behaviour $(m-1,1)$-type correlator when $A$ and $B$ are far away, therefore it is an IR constant. On the other hand, $C[h_i]$ characterizes the logarithmic divergent behaviour of $(m)$-type correlator when $A$ and $B$ coincide, so it is a UV constant. In this sense, \eqref{cdrel} is a UV/IR relation. 
\end{enumerate}
We will comment on the analytic continuation of conformal block further at the end of this section. There are different ways to seperate $m$ OPE blocks, the general way is to uplift $(m)$-type correlator to $(p,m-p)$-type, 
\be
\langle Q_A[\mathcal{O}_1]\cdots Q_A[\mathcal{O}_m]\rangle_c\xrightarrow{split}\langle Q_A[\mathcal{O}_1]\cdots Q_A[\mathcal{O}_p]Q_B[\mathcal{O}_{p+1}]\cdots Q_B[\mathcal{O}_m]\rangle_c,\quad 1\le p\le m-1.
\ee 
As has been noticed, the right hand side only depends on cross ratio $\eta$ and finite, though it is not necessary conformal block. Nevertheless, we should find 
\be
\langle Q_A[\mathcal{O}_1]\cdots Q_A[\mathcal{O}_m]\rangle_c=\lim_{\eta\to-1}\langle Q_A[\mathcal{O}_p]Q_B[\mathcal{O}_{p+1}]\cdots Q_B[\mathcal{O}_m]\rangle_c.\label{mmp}
\ee 
This may put contraints on possible structure of $(p,m-p)$-type correlator since $(m)$-type is only logarithmically divergent. We can test \eqref{mmp} using the $(2,2)$-type correlator of modular Hamiltonians of \cite{Long:2019fay}, we read 
\bea
\lim_{B\to A}\langle H_A^2H_B^2\rangle_c&=&\lim_{\eta\to-1}
c\{\frac{1+\eta}{\eta^{2}}[4 \mathrm{Li}_{3}(1+\eta)-2 \log (1+\eta) \mathrm{Li}_{2}(1+\eta)+\frac{2 \log (1+\eta)}{3} \operatorname{Li}_{2}(-\eta)\nn\\&&+\frac{1+\eta}{3} \log ^{2}(1+\eta)
-\frac{\pi^{2}}{3} \log (1+\eta)-4 \zeta(3)]+\frac{2+\eta}{3 \eta}\left[2 \operatorname{Li}_{2}(-\eta)+3 \log (1+\eta)\right]-\frac{4}{3}\}\nn\\&\sim&2c\log\frac{2}{\epsilon},\quad \epsilon\to0.
\eea 
This is consistent with \eqref{x2222}. We can extend this observation further since one can uplift $(m)$-type correlator to $(m_1,\cdots,m_k)$-type correlator with $\sum_{i=1}^k m_i=m$. We hope to return to this topic in the future. 

\section{More logarithmic behaviour}
The logarithmic behaviour in Section \ref{loglaw} happens when region $B$ coincide with region $A$. When region $B$ attaches to region A, $(m,n)$-type correlator may also be divergent. To check this point,  
 we parameterize the two intervals as 
\be
x_1=1,\quad x_2=-1,\quad x_3=3+\epsilon,\quad x_4=1+\epsilon, \quad \epsilon\to 0.
\ee 
The cross ratio 
\be
\eta=\frac{1}{\epsilon}+\cdots
\ee 
approches infinity when $x_3\to x_2$. 
Then conformal block 
\be
G_h(\eta)=\eta^h \ {}_2F_1(h,h,2h,-\eta)\sim \frac{2^{2h-1}\Gamma(h+\frac{1}{2})}{\sqrt{\pi}\Gamma(h)}\log\frac{1}{\epsilon}+\cdots,\quad \epsilon\to 0.
\ee 
We use the notation 
\be
\langle Q_A[\mathcal{O}_1]\cdots Q_A[\mathcal{O}_m]\odot Q_B[\mathcal{O}_{m+1}]\cdots Q_B[\mathcal{O}_{m+n}]\rangle_c
\ee to denote the corresponding correlator, then
\be
\langle Q_A[\mathcal{O}_1]\cdots Q_A[\mathcal{O}_m]\odot Q_B[\mathcal{O}]\rangle_c=C'[h_1,\cdots,h_m,h]\log\frac{1}{\epsilon}
\label{Cp}
\ee 
is divergent logarithmically when the two intervals attach to each other. This is another type of logarithmic behaviour. Interestingly, the coefficient 
\be
C'[h_1,\cdots,h_m,h]=\frac{2^{2h-1}\Gamma(h+\frac{1}{2})}{\sqrt{\pi}\Gamma(h)}D[h_1,\cdots,h_m,h]=\frac{(-1)^{-h}}{2}C[h_1,\cdots,h_m,h].
\ee is not  independent. One can also consider general $(m_1,m_2,\cdots,m_{k})$-type correlator, for example, $(2,2)$-type correlator 
\be
\langle H_A^2\odot H_B^2\rangle_c\sim \  c\ \log\eta,\quad \eta=\frac{1}{\epsilon}\to\infty.
\ee
However, it is unclear whether these correlators always have logarithmic behaviour.

\section{Deformed reduced density matrix and grand canonical ensemble of a subsystem}
It has been noticed that connected correlation function of OPE blocks \cite{Long:2019fay, Long:2019pcv} can be generated by the expectation value of deformed reduced density matrix 
\be
\rho_A=e^{-W},\quad W=\sum_i \alpha_i Q_A[\mathcal{O}_i]\label{def}.
\ee 
More explicitly, they are generated by 
\be
\log Z(\alpha_i)=\log \langle\rho_A\rangle=\log\text{tr}_A e^{-H_A-W}.
\ee 
The function $Z(\alpha_i)$ is similar to the partition function (grand potential) in equilibrium thermodynamic systems. For example, assume
\be
W=a H_A+\alpha Q_A
\ee 
where $Q_A$ is an OPE block, then 
\be
[H_A,Q_A]=0.
\ee 
Reparameterizing
\be
\beta=1+a,\quad \alpha=-\beta\  \mu,
\ee 
therefore
\be
Z(\alpha_i)=Z(\beta,\mu)=\text{tr}_A e^{-\beta(H_A-\mu Q_A)}=e^{-\beta \Omega_A(\beta,\mu)},
\ee 
where 
\be
\Omega_A(\beta,\mu)=-\beta^{-1}\log \text{tr}_A e^{-\beta(H_A-\mu Q_A)}=-\beta^{-1}\log\langle\rho_A\rangle
\ee 
is defined to be ``grand potential'' of subsystem.  The parameter $\beta$ is inverse ``temperature''
\be
\beta=\frac{1}{T}.
\ee 
and $\mu$ is ``chemical potential'' associated with OPE block $Q_A$. ``Gibbs entropy'' is 
\be
S_A(\beta,\mu)=-\frac{\partial \Omega_A(\beta,\mu)}{\partial T}|_{\mu\  \text{fixed}}.
\ee
One can easily show that ``Gibbs entropy'' is von Neumann entropy 
\be
S_A(\beta,\mu)=-\text{tr}_A\tilde{\rho}_A\log\tilde{\rho}_A
\ee 
where $\tilde{\rho}_A$ is normalized 
\be
\tilde{\rho}_A=\frac{e^{-\beta(H_A-\mu Q_A)}}{\text{tr}_A e^{-\beta(H_A-\mu Q_A)}},\quad \text{tr}_A\tilde{\rho}_A=1.
\ee 
``Gibbs entropy'' defined in this way is not usual entanglement entropy in general. 
Expectation value of ``energy'' $H_A$ and ``charge'' $Q_A$ are 
\be
\mathcal{E}_A\equiv\text{tr}_A \tilde{\rho}_A H_A,\quad \mathcal{Q}_A\equiv \text{tr}_A\tilde{\rho}_A Q_A.
\ee 
Then ``grand potential'', ``Gibbs entropy'', ``energy'' and ``charge'' satisfy the identity 
\be
\mathcal{E}_A=\Omega_A+\mu \mathcal{Q}_A+T S_A.\label{identitye}
\ee 
Note ``charge'' can also be generated by ``grand potential''
\be
\mathcal{Q}_A=-\frac{\partial \Omega_A(\beta,\mu)}{\partial\mu}|_{\beta\ \text{fixed}},
\ee 
differential of ``grand potential'' becomes 
\be
d\Omega_A=-S_AdT-\mathcal{Q}_A d\mu.\label{diff}
\ee 
Combining \eqref{identitye} and \eqref{diff}, we obtain an equation of ``first law of theromodynamics'' for any subsystem 
\be
d\mathcal{E}_A=TdS_A+\mu d\mathcal{Q}_A.\label{firstlaw}
\ee 
The ``first law'' should be distinguished with ``first law of entanglement'' found in \cite{Blanco:2013joa}. The former is valid at all order while the latter is valid up to linear order of perturbation. 
 
The direct consequence of \eqref{corre} is that any deformed reduced density matrix \footnote{It would be interesting to study the case when OPE blocks $Q_A[\mathcal{O}_i]$ are non-commutative with each other.} may be associated with a logrithmic behaviour 
\be
\log\langle \rho_A\rangle=F(\alpha_i)\log\frac{L}{\epsilon}
\ee 
in CFT$_2$. The function $F(\alpha_i)$ is a generator of coefficients $C[h_i]$. 
It is obtained in \eqref{TAa} where the operator is quadratic for free scalar theory. In previous example, $W=a H_A+\alpha Q_A$, we find 
\bea
\Omega_A(\beta,\mu)&=&-\beta^{-1}F(\beta,\mu)\log\frac{L}{\epsilon},\\
S_A(\beta,\mu)&=&(1-\beta\frac{\partial}{\partial\beta})F(\beta,\mu)\log\frac{L}{\epsilon},\\
\mathcal{Q}_A&=&\beta^{-1}\frac{\partial}{\partial\mu}F(\beta,\mu)\log\frac{L}{\epsilon},\\
\mathcal{E}_A&=&\frac{1}{\beta}(\mu\frac{\partial}{\partial\mu}-\beta\frac{\partial}{\partial\beta})F(\beta,\mu)\log\frac{L}{\epsilon}.
\eea 

\section{Area law in higher dimensions}
In this section, we will briefly comment on connected correlation function of OPE blocks to higher dimensions. We assume region $A$ and $B$ are disjoint, their radius are $R$ and $R'$, respectively. The distance of their center is $r$, $r>R+R'$. We assume they are located at the same time slice, therefore there is only one independent cross ratio 
\be
z=x=\frac{4RR'}{r^2-(R-R')^2}.
\ee  
OPE block corresponding to a primary operator $\mathcal{O}$ of dimension $\Delta$ and spin $J$ is \cite{deBoer:2016pqk}
\be
Q_A[\mathcal{O}]=c_{\Delta,J}\int_{\mathcal{D}(A)} d^dx K^{\mu_1}\cdots K^{\mu_J}|K|^{\Delta-J-d}\mathcal{O}_{\mu_1\cdots \mu_J},
\ee 
where vector $K^{\mu}$ is the generator of modular Hamiltonian of spherical region of CFT$_d$. Constant $c_{\Delta,J}$ depends on conventions and we  leave it free. The integral region is in the domain of dependence of $A$. When the operator is conserved, the integral can be reduced to an integral inside region $A$. One special example is modular Hamiltonian $H_A$ itself. One can construct $(m,n)$-type correlators 
\be
\langle Q_A[\mathcal{O}_1]\cdots Q_A[\mathcal{O}_m]Q_B[\mathcal{O}_{m+1}]\cdots Q_B[\mathcal{O}_{m+n}]\rangle_c,\quad m,n\ge 1.
\ee 
As $d=2$ case, the correlators are finite in general. For $n=1$, $(m,1)$-type correlator may be a conformal block 
\be
\langle Q_A[\mathcal{O}_1]\cdots Q_A[\mathcal{O}_m]Q_B[\mathcal{O}]\rangle_c=D[\Delta_i]G_{\Delta,J}(z),\label{corrh}
\ee 
where $G_{\Delta,J}(z)$ is a conformal block in general $d$ dimensions. The reason is the same as two dimensions. The structure \eqref{corrh} are guaranteed  by conformal symmetry.  $Q_B[\mathcal{O}]$ is an eigenvector under the action of Casimir operator of conformal algebra, combining with boudary conditions when region $A$ shrinks to a point, $(m,1)$-type correlators should be a conformal block. Now the conformal block can also be analytically continued to the limit 
\be
r\to0,\quad R'\to R.
\ee 
Just as $d=2$ case, we continue the function by perserving spherical symmetry, 
\be
r=0,\quad  R=1,\quad R'=1-\epsilon,\quad \epsilon\to 0.
\ee 
The cross ratio approaches $-\infty$ by  
\be
z=-\frac{4(1-\epsilon)}{\epsilon^2},\quad \epsilon\to0.
\ee 
Now it is interesting to check the asymptotic behavaiour of conformal block in this limit. The general $d$ dimensional conformal block in the diagonal limit ($z=x$)can be found in \cite{Hogervorst:2013kva}. For any finite conformal weight $\Delta$ and spin $J$, it is a finite sum of ${}_3F_2$ functions. 
 We collect the discussion on conformal block in Appendix \ref{appc}. Hypergeometric function behaves as
\bea
&&{}_3F_2(\alpha,\beta,\gamma;\xi,\zeta;Z)\nn\\&\sim& \frac{\Gamma(\xi)\Gamma(\zeta)}{\Gamma(\alpha)\Gamma(\beta)\Gamma(\gamma)}[\frac{\Gamma(\alpha)\Gamma(\beta-\alpha)\Gamma(\gamma-\alpha)}{\Gamma(\zeta-\alpha)\Gamma(\xi-\alpha)}((-Z)^{-\alpha}+\cdots)\nn\\&&+\frac{\Gamma(\beta)\Gamma(\alpha-\beta)\Gamma(\gamma-\beta)}{\Gamma(\zeta-\beta)\Gamma(\xi-\beta)}((-Z)^{-\beta}+\cdots)+\frac{\Gamma(\gamma)\Gamma(\alpha-\gamma)\Gamma(\beta-\gamma)}{\Gamma(\zeta-\gamma)\Gamma(\xi-\gamma)}((-Z)^{-{\gamma}}+\cdots)],\nn\\
\eea 
where 
\be
Z=\frac{z^2}{4(z-1)}=-\frac{4(1-\epsilon)^2}{\epsilon^2(2-\epsilon)^2}=-\frac{1}{\epsilon^2}+\cdots.\label{Z}
\ee 
Conformal block is divergent 
\be
G_{\Delta,J}(z)\sim \text{const.}\times (-Z)^{\upsilon},\quad Z\to-\infty
\ee 
where 
\be
\upsilon=\frac{d}{2}-1,\quad d>2.
\ee 
The $\text{const.}$ term  can be obtained directly, however, we don't need them in this paper. $(m)$-type correlator becomes \footnote{To simplify notation, we will consider the correlator of identical OPE blocks.}
\be
\langle Q_A[\mathcal{O}]^m\rangle_c=C[\Delta,J](\frac{R}{\epsilon})^{d-2}+\cdots.
\ee 
We have inserted back the radius $R$. Terms in $\cdots$ are subleading, we will discuss them later. Note the area of the boundary of $A$ is 
\be
\mathcal{A}\sim R^{d-2},\quad d>2,
\ee 
we conclude that $(m)$-type obeys area law 
\be
\langle Q_A[\mathcal{O}]^m\rangle_c=C[\Delta,J]\frac{\mathcal{A}}{\epsilon^{d-2}}+\cdots.\label{arealaw}
\ee 
When the OPE block is modular Hamiltonian, \eqref{arealaw} is equivalent to area law of R\'enyi entropy \cite{Srednicki:1993im}. In R\'enyi entropy, one may read cutoff independent information from subleading terms. For example, for a system with gravitational dual, the typical behaviour of entanglement entropy is \cite{Ryu:2006bv}
\bea
S_{A}\sim \left\{\begin{aligned}
&p_1 (\frac{R}{\epsilon})^{d-2}+p_2 (\frac{R}{\epsilon})^{d-4}+\cdots +q_0+\cdots,\quad &\text{d odd},\\&p_1 (\frac{R}{\epsilon})^{d-2}+p_2 (\frac{R}{\epsilon})^{d-4}+\cdots +q\log\frac{R}{\epsilon}+\cdots,\quad &\text{d even}.
\end{aligned}\right.\label{holo}
\eea 
$q$ is cutoff independent and encodes useful information of the theory. In a similar way, we would like to understand the subleading behaviour in $(m)$-type correlator. It turns out that the subleading behaviour depends on dimension $\Delta$ and $J$. We will check several examples in the following. 
\begin{enumerate}
	\item $d=3, J\le10$. The general behaviour is 
	\be
	G_{\Delta,J}(z)\sim \gamma \frac{\mathcal{A}}{\epsilon'}+q_1 \log\frac{R}{\epsilon'}+q_0 +\cdots
	\ee 
	where $\epsilon'$ is not necessary to be $\epsilon$ in \eqref{Z}, but in the small $\epsilon$ limit, they are of the same order 
	\be
	\epsilon'=\epsilon+o(\epsilon). \label{ep}
	\ee 
	For general conformal weight, $q_1\not=0$, therefore $q_1$ is cutoff independent. Only for special operators, for example, stress tensor has  $J=2,\Delta=3$, $q_1=0$, then in this case, $q_0$ is cutoff independent. This is consistent with holographic result \eqref{holo}. Interestingly, when the operator is a symmtric traceless conserved current, its dimension obeys unitary bound \cite{Minwalla:1997ka}
	\be
	\Delta=J+1,\quad \text{for}\quad J\ge1,
	\ee 
	we always find $q_1=0$. 
	\item $d=4,J\le10$. The general behaviour is 
	\be
	G_{\Delta,J}(z)\sim \gamma \frac{\mathcal{A}}{\epsilon'^2}+q_2\log^2\frac{R}{\epsilon'}+q_1 \log\frac{R}{\epsilon'}+q_0+\cdots \label{4dcb}
	\ee 
	Again, 
	$\epsilon'$ is \eqref{ep}, it can be chosen such that no linear term $\frac{R}{\epsilon'}$ in \eqref{4dcb}.
	In general, $q_2$ is cutoff independent. One simple example is $\Delta=4,J=0$, 
	\be
	G_{4,0}(z)=-48(Z-\arcsin^2\sqrt{Z})^2
	\ee  However, for special combination of $\Delta$ and $J$, $q_2=0$. One example is stress tensor $\Delta=4,J=2$, it has $q_2=0,q_1\not=0$, therefore $q_1$ is cutoff independent. Again, this is consistent with holographic result \cite{Ryu:2006bv}. We also observe that $q_2=0$ for 
\be
\Delta=J+2,\quad J\ge1,
\ee 
which is exactly unitary bound for a symmetric traceless conserved current in four dimensions.
\item In  general higher dimensions, one may find
\bea
G_{\Delta,J}(z)\sim \left\{\begin{aligned}&\gamma\ \frac{\mathcal{A}}{\epsilon'^{d-2}}+p_{d-4}(\frac{R}{\epsilon'})^{d-4}+\cdots+q_1\log\frac{R}{\epsilon'}+q_0+\cdots\quad&\text{d odd},\\&\gamma\ \frac{\mathcal{A}}{\epsilon'^{d-2}}+p_{d-4}(\frac{R}{\epsilon'})^{d-4}+\cdots+q_2\log^2\frac{R}{\epsilon'}+q_1\log\frac{R}{\epsilon'}+q_0+\cdots,\quad&\text{d even}.\end{aligned}\right.\label{gecb}
\eea
For general conformal weight $\Delta$ and spin $J$, the logarithmic behaviour  in \eqref{gecb}  is not the same as \eqref{holo}. When the operator is a conserved current 
\be
\Delta=J+d-2,\quad J\ge1,
\ee
it recovers the same phenomenon \eqref{holo}.  

\end{enumerate}

We can also consider the limit $B$ attaches to $A$, 
\be
R=1,\quad R'=1,\quad r=2+\epsilon,\quad \epsilon\to 0^+.
\ee 
The cross ratio approaches $1^-$ 
\be
z=\frac{4}{(2+\epsilon)^2}=1-\epsilon+\cdots.
\ee 
The variable $Z$ still approaches $-\infty$ 
\be
Z=-\frac{4}{\epsilon(4+\epsilon)(2+\epsilon)^2}=-\frac{1}{4\epsilon}+\cdots,
\ee 
though much more slower than \eqref{Z}. Therefore we find a new `` law'' of divergence for connected correlation function 
\bea
\langle Q_A^{m-1}\odot Q_B\rangle_c\sim \left\{\begin{aligned}&\nu \frac{\mathcal{L}}{\epsilon'^{\frac{d-2}{2}}}+p_{d-4}(\frac{R}{\epsilon'})^{\frac{d-4}{2}}+\cdots+q_1\log\frac{R}{\epsilon'}+q_0+\cdots
,\quad& \text{d odd}\\&\nu \frac{\mathcal{L}}{\epsilon'^{\frac{d-2}{2}}}+p_{d-4}(\frac{R}{\epsilon'})^{\frac{d-4}{2}}+\cdots+q_2\log^2\frac{R}{\epsilon'}+q_1\log\frac{R}{\epsilon'}+q_0+\cdots,\quad &\text{d even}.\end{aligned}\right.\eea 
We use $\mathcal{L}$ to denote the square root of area
\be
\mathcal{L}=\sqrt{\mathcal{A}}.
\ee 
in four dimensions, $\mathcal{L}$ has the dimension of length. We also find $q_1=0$ in odd dimensions and $q_2=0$ in even dimensions for OPE block associated with conserved current. 

\section{Conclusion and discussion}
Connected correlation functions of $(m)$-type in CFT$_2$ have been studied extensively.  They are divergent and obey logarithmic law, which is summarized by formula \eqref{corre}. The coefficient $C[h_i]$ is independent of UV cutoff and encodes details of CFT$_2$. We could check this formula by direct regularization of $(m)$-type correlators. To justify our results, we also provide three different ways to deal with $(m)$-type correlators. Firstly, we calculate the logarithm of the expectation value of an exponential operator $e^{-H_A-\sum_{s\ge 2}\alpha_s Q_A[J_s]}$, the result is presented in \eqref{TAa}. Secondly, we use Selberg integral to regularize several correlators, we could also find logarithmic behaviour by carefully identifying the regularization parameter. Finally, we use analytical continuation of $(p,m-p)$-type correlators to reproduce the same results. Interestingly, $(p,m-p)$-type correlation functions considered in \cite{Long:2019fay,Long:2019pcv} are finite when the two regions $A$ and $B$ are disjoint. They admit continuation to the case when $A$ and $B$ coincide. The uplift from $(m)$-type to $(p,m-p)$-type is arbitrary, but the final coefficients $C[h_i]$ should be unique.  Since $(m-1,1)$-type correlator is conformal block  up to a constant $D[h_i]$, any $(m)$-type correlator may have logarithmic divergence, the coefficient $C[h_i]$ is related to $D[h_i]$ by identity \eqref{cdrel}. 

Motivated by the powerful method of analytic continuation of conformal block, we find new logarithmic behaviour when region $B$ attaches to region $A$. This is shown in formula \eqref{Cp}.  A novel property is that the three coefficients 
\be
D[h_i],\quad C[h_i]\quad\text{and}\quad C'[h_i]
\ee 
are not independent. Mathematically, this follows from the asymptotic behaviour of conformal block near
\be
\eta=0,\quad -1\quad \text{and} \quad \infty,
\ee 
respectively. Physically, it is a rather intriguing UV/IR relation in connected correlation functions. 

There are many problems to be solved in this direction. 
\begin{enumerate}
	\item Deformed reduced density matrix \eqref{def} is claimed to be meaningful in \cite{Long:2019fay, Long:2019pcv} by its formal similarity to Wilson loop \cite{Maldacena:1998im, Rey:1998ik}. In this paper, its expectation value $\langle \rho_A\rangle=\text{tr}_A\  e^{-H_A-W}$ is shown to be controllable, in the sense that it obeys logarithmic law. We notice that it is analogous to grand canonical ensemble in equilibrium thermodynamic system. We define corresponding ``grand potential'', ``Gibbs entropy'', `` energy'' and `` charge'' for a subsystem. A ``first law of thermodynamics'' for a subsystem has been derived in parallel to statistic mechanics. However, the physical meaning of these quantities deserves further study.
	\item We discussed the area law of $(m)$-type correlators in higher dimensions from analytic continuation of conformal block. Interestingly, this is equivalent to the area law associated with exponential of OPE block
	\bea
	\log\langle e^{-\alpha Q_A[\mathcal{O}]}\rangle\sim \left\{\begin{aligned}&\gamma\ \frac{\mathcal{A}}{\epsilon^{d-2}}+p_{d-4}(\frac{R}{\epsilon})^{d-4}+\cdots+q_1\log\frac{R}{\epsilon}+q_0+\cdots\quad&\text{d odd},\\&\gamma\ \frac{\mathcal{A}}{\epsilon^{d-2}}+p_{d-4}(\frac{R}{\epsilon})^{d-4}+\cdots+q_2\log^2\frac{R}{\epsilon}+q_1\log\frac{R}{\epsilon}+q_0+\cdots,\quad&\text{d even}.\end{aligned}\right.\label{gecba}
	\eea
	For conserved current $\mathcal{J}$, the area law becomes 
		\bea
	\log\langle e^{-\alpha Q_A[\mathcal{J}]}\rangle\sim \left\{\begin{aligned}&\gamma\ \frac{\mathcal{A}}{\epsilon^{d-2}}+p_{d-4}(\frac{R}{\epsilon})^{d-4}+\cdots+q_0+\cdots\quad&\text{d odd},\\&\gamma\ \frac{\mathcal{A}}{\epsilon^{d-2}}+p_{d-4}(\frac{R}{\epsilon})^{d-4}+\cdots+q_1\log\frac{R}{\epsilon}+q_0+\cdots,\quad&\text{d even}.\end{aligned}\right.\label{gecba2}
	\eea
	This is a direct generalization of R\'enyi entanglement entropy \eqref{holo}. It is better to validate this behaviour in more details. The cutoff independent coefficients before $\log\frac{R}{\epsilon}$ or $\log^2\frac{R}{\epsilon}$ encode rich CFT data, there is no work on how to find out them so far. In this paper, we find one, $F(\alpha_i)$, for $2d$ massless free scalar theory. 

	\item The logarithmic behaviour in higher dimensions \eqref{gecba2} can be continued to $d=2$, which is consistent with \eqref{2dlog}. Actually, for chiral primary operators $\mathcal{O}(z)$ considered in the context, it is conserved in the sense 
	\be
	\bar{\partial}\mathcal{O}(z)=0. 
	\ee 
	Its conformal weight $\Delta=h$ and spin $J$ are equal, therefore they satisfy the unitary bound 
	\be
	\Delta=J=h,\quad d=2.
	\ee 
	In two dimensions, there are non-chiral primary operators $\mathcal{O}(z,\bar{z})$ whose conformal weight and spin are 
	\be
	\Delta=h+\bar{h}, J=h-\bar{h}.
	\ee 
	The conformal weight is larger than spin in general. 
	The OPE block for these operators are 
	\be
	Q_A[\mathcal{O}]=c_{h,\bar{h}}\int_{z_1}^{z_2}dz\int_{\bar{z}_1}^{\bar{z}_2}d\bar{z} \left(\frac{(z-z_1)(z_2-z)}{(z_2-z_1)}\right)^{h-1}\left(\frac{(\bar{z}-\bar{z}_1)(\bar{z}_2-\bar{z})}{\bar{z}_2-\bar{z}_1}\right)^{\bar{h}-1}\mathcal{O}(z,\bar{z}).
	\ee 
	They should obey another logarithmic law 
	\be
	\log \langle e^{-\alpha Q_A[\mathcal{O}]}\rangle=q_2\log^2\frac{L}{\epsilon}+q_1\log\frac{L}{\epsilon}+q_0+\cdots,
	\ee
	which is consistent with analytic continuation from \eqref{gecba} in higher dimensions to $2d$. One can also understand the $\log^2\frac{L}{\epsilon}$ divergence in another way. The holomorphic part contributes one $\log\frac{L}{\epsilon}$ and anti-holomorphic part contributes another logarithmic divergence. Therefore, the divergent behaviour is the product of holomorphic and anti-holomorphic part, which is exactly $\log^2\frac{L}{\epsilon}$.

	\item We defined another type of correlators when region $A$ attaches region $B$
	\be
	\langle Q_A[\mathcal{O}_1]\cdots Q_A[\mathcal{O}_m]\odot Q_B[\mathcal{O}_{m+1}]\cdots Q_B[\mathcal{O}_{m+n}]\rangle_c.
	\ee 
	These correlators are also divergent. However,
	we showed in several examples that one can also obtain cutoff independent information from the divergent behaviour. For $m,n\ge 2$, we can not use the trick of analytic continuation of conformal block. Therefore it is not guaranteed that it  always obeys logrithm law in two dimensions (or area law in higher dimensions). One should develop other method to tackle this problem.  In higher dimensions, a new divergent behaviour for $\langle Q_A^{m-1}\odot Q_B\rangle_c$ is also very interesting.

	\item In AdS/CFT correspondence, one could extract OPE data from holographic correlation function using Witten diagrams \cite{Witten:1998qj}. ``Geodesic Witten diagram'' \cite{Hijano:2015rla, Hijano:2015zsa} is a bulk description of conformal block. On the other hand, $(m,1)$-type correlators are also claimed to be conformal block. It would be better to establish a  relationship between geodesic Witten diagram and $(m,1)$-type correlator. OPE block has a natural gravitational dual in the bulk, for example,  the dual object in large $N$ limit is schematically 
	\be
	\int_{\gamma}\phi ,
	\ee 
	where $\gamma$ is minimal surface associated with boundary region $A$, $\phi$ is the bulk field which is dual to primary operator $\mathcal{O}$. Our work suggests that the following exponential operator 
	\be
   W_{\gamma}(\phi)=	e^{\int_{\gamma}\phi}
	\ee 
	may encode rich holographic CFT data. To extract holographic data, one can compute the logarithm of its expectation value
	\be
	\log\langle W_{\gamma}(\phi)\rangle.
	\ee 
	We expect that it obeys area law in higher dimensions and logarithmic law in two dimensions. For two disjoint regions, one may consider the following correlation functions 
	\be
	I_{\gamma_1,\gamma_2}(\phi_1,\phi_2)=\log \langle W_{\gamma_1}(\phi_1) W_{\gamma_2}(\phi_2)\rangle-\log\langle W_{\gamma_1}(\phi_1)\rangle-\log\langle W_{\gamma_2}(\phi_2)\rangle,
	\ee 
	where $\gamma_i, \ i=1,2$ is minimal surface corresponds to region $A$ and $B$. The CFT results in \cite{Long:2019fay, Long:2019pcv} indicate that $I_{\gamma_1,\gamma_2}(\phi_1,\phi_2)$ is finite in large $N$ limit.
	\item We introduced an integral \eqref{SNG}  with $\frac{n(n+3)}{2}$ parameters to deal with analytic continuation of integrals in this paper. This integral is crucial for our understanding of general connected correlation functions. When $n=1$, the integral is an integral representation of Beta function. When $n=2$ and $\alpha_1=\alpha_2,\ \beta_1=\beta_2$, the function is standard Selberg integral. Unfortunately, a closed formula is lack for general cases, as far as we know. This integral may reveal deep connection between mathematics and physics.
\end{enumerate}


\vspace{5mm}

\begin{center}
\textbf{\Large{	Appendices}}
\end{center}
\appendix
\section{$(m,1)$-type correlator}\label{AA}
In this section, we show that $(m,1)$-type correlator is conformal block \eqref{conf}. We use the differential method as \cite{Dolan:2003hv}. Assume the end points of $A$ are $z_1,z_2$ and  $z_3,z_4$ for $B$, then since OPE blocks in this paper are invariant under conformal transformation, a general $(m,1)$-type correlator is only a function of cross ratio 
\be
\langle Q_A[\mathcal{O}_1]\cdots Q_A[\mathcal{O}_m]Q_B[\mathcal{O}]\rangle_c=H(\eta),\quad \eta=\frac{z_{12}z_{34}}{z_{14}z_{23}}. 
\ee 
Casimir operator of $SL(2,R)$ is 
\be
\mathcal{L}_2=\frac{1}{2}L_{AB}L^{AB}
\ee 
where $L_{AB}$ are generators of global $SL(2,R)$. It is a summation $SL(2,R)$ generators act on each end point of $B$, 
\be
L_{AB}=L^{(z_3)}_{AB}+L^{(z_4)}_{AB}.
\ee 
In coordinate representation, the holomorphic part of Casimir $\mathcal{L}_2$ is 
\be
\mathcal{L}_2=z_{34}^2\partial_{z_3}\partial_{z_4}.
\ee  
Casimir operator acts on $(m,1)$-type correlator, then  
\bea
\mathcal{L}_2 \langle Q_A[\mathcal{O}_1]\cdots Q_A[\mathcal{O}_m]Q_B[\mathcal{O}]\rangle_c&=&z_{34}^2\partial_{z_3}\partial_{z_4} \langle Q_A[\mathcal{O}_1]\cdots Q_A[\mathcal{O}_m]Q_B[\mathcal{O}]\rangle_c\nn\\
&=&z_{34}^2\partial_{z_3}\partial_{z_4} H(\eta)\nn\\&=&-\eta^2(1+\eta)H''-\eta^2 H'.
\eea 
On the other hand, OPE block $Q_B[\mathcal{O}]$ is an eigenvector of Casimir operator with eigenvalue $-h(h-1)$, 
\be
[\mathcal{L}_2, Q_B[\mathcal{O}]]=-h(h-1)Q_B[\mathcal{O}].
\ee 
Therefore, 
\be
\eta^2(1+\eta)H''+\eta^2 H'=h(h-1)H.\label{H}
\ee 
Note the leading order of $H(\eta)$ is $\eta^h$ since 
\be
Q_B[\mathcal{O}]\xrightarrow{z_3\to z_4} (z_3-z_4)^h \mathcal{O}.
\ee 
Therefore the solution of \eqref{H} is 
\be
H(\eta)=D[h_i,h]\ \eta^h \ {}_2F_1(h,h,2h,-\eta)
\ee 
which is exactly conformal block. Since Casimir operator is a second-order partial differential operator, $(m,n)$-type correlator with $n\ge2$ is not conformal block in general. 

\section{Integral}
The basic integral ${}_jI_4[h,h,h,h]$ can be regularized as $I_2[h]$. There is one subtlety when $j=2h$ since this integral shows mixed divergent behaviour 
\be
{}_{2h}I_4[h,h,h,h]=b_4[h,2h]\log\frac{2}{\epsilon}+b_4[h,2h]\log^2\frac{2}{\epsilon}.
\ee 
One should carefully seperate these two classes of divergent behaviour. We will use $h=1$ as an example to show how to calculate such kind of integrals. 
\bea
{}_2I_4[1,1,1,1]&=&\prod_{i=1}^4 \int_{-1}^1dz_i \frac{\theta^2}{z_{12}^2z_{34}^2}\nn\\&=&\int_{-1+\epsilon}^{1-\epsilon} dz_3\int_{-1+\epsilon}^{1-\epsilon} dz_4[ \frac{8}{(1-z_3^2)(1-z_4^2)}-\frac{8(\text{arctanh}z_3-\text{arctanh}z_4)^2}{(z_3-z_4)^2}]
\eea 
At the second step, we use integrate $z_1,z_2$ by ignoring any poles in the integrand. We insert UV cutoff to regularize the integral. Note $b_4[1,2]$ should be $2b_4[1,0]$
\be
b_4[1,2]=2b_4[1,0]=8
\ee 
since the $\log^2\frac{2}{\epsilon}$ terms should be canceled in the connected correlation function. Therefore we may subtract a term
\be
8\times( \int_{-1+\epsilon}^{1-\epsilon} \frac{1}{1-z_3^2})^2
\ee in the integral. 
\be
{}_2I_4[2,2,2,2]=8\log^2\frac{2}{\epsilon}-\int_{-1+\epsilon}^{1-\epsilon}dz_3 \int_{-1+\epsilon}^{1-\epsilon}dz_4 \frac{8(\text{arctanh}z_3-\text{arctanh}z_4)^2}{(z_3-z_4)^2}
\ee 
Now we change variables to $y_3,y_4$ by
\be
z_3=\tanh\frac{y_3}{2},\quad z_4=\tanh\frac{y_4}{2}.
\ee 
Then 
\bea
{}_2I_4[1,1,1,1]&=&=8\log^2\frac{2}{\epsilon}-\int_{-\log\frac{2}{\epsilon}}^{\log\frac{2}{\epsilon}} dy_3 \int_{-\log\frac{2}{\epsilon}}^{\log\frac{2}{\epsilon}} dy_4 \frac{(y_3-y_4)^2}{2\cosh^2\frac{y_3-y_4}{2}}\nn\\
&=&8\log^2\frac{2}{\epsilon}-\frac{8\pi^2}{3}\log\frac{2}{\epsilon}.
\eea 
Therefore, we find a way to seperate $\log\frac{2}{\epsilon}$ from ${}_{2h}I_4[h,h,h,h]$
\be
{}_{2h}I_4[h,h,h,h]=2b_4'[h,0](\int_{-1+\epsilon}^{1-\epsilon}dz_1 \frac{1}{1-z_1^2})^2+{}_{2h}\tilde{I}_4[h,h,h,h],
\ee
where 
\be
{}_{2h}\tilde{I}_4[h,h,h,h]=\prod_{i=1}^4 \int_{-1}^1 dz_i(1-z_i^2)^{h-1} \frac{\theta^{2h}-2}{z_{12}^{2h}z_{34}^{2h}}.
\ee 
The integral ${}_{2h}I_4[h,h,h,h]$ turns out to be logarithmically divergent. 
\section{Analytic continuation}\label{analy}
In Section \ref{analycb}, we obtain logarithmic behaviour by taking the limit $B\to A$. We choose the way to approach the coincidence limit symmetrically, 
\be
x_1=1,\quad x_2=-1,\quad x_3=1-\epsilon,\quad x_4=-1+\epsilon.
\ee
The center of two intervals are the same while the radius are not. 
 However, since two regions are non-local, there are other ways. We could choose the center of two intervals are not coincide, 
 \be
 x_1=1,\quad x_2=-1,\quad x_3=1+\epsilon,\quad x_4=-1+\epsilon.
 \ee 
 Then the cross ratio 
 \be
 \eta=\frac{4}{-4+\epsilon^2}=-1-\frac{\epsilon^2}{4}+\cdots,
 \ee 
therefore the conformal block is 
\be
G_h(\eta)=\frac{2(-1)^h \Gamma(2h)}{\Gamma(h)^2}\log\frac{2}{\epsilon}+\frac{2(-1)^{h+1}\Gamma(2h)(i\pi+\gamma_E+\psi(h))}{\Gamma(h)^2}+\cdots.
\ee 
$\gamma_E$ is Euler Gamma constant 
\be
\gamma_E\approx 0.577216
\ee 
which is the limit 
\be
\gamma_E=\lim_{n\to\infty}(\sum_{k=1}^n\frac{1}{k}-\frac{1}{n}).
\ee 
 $\psi(z)$ is the derivative of logarithm of Gamma function 
\be
\psi(z)=\frac{\Gamma(z)'}{\Gamma(z)}.
\ee 
The logarithmic behaviour is the same as \eqref{cbex}, including the coefficient. The problem appears in the constant term, which is complex due to the term proportional to $i\pi$. However, it doesn't affect the result in this paper.

\section{Conformal block in general dimensions}\label{appc}
We will collect some basic facts about conformal block in general dimensional conformal field theory. The reader is  referred to \cite{Dolan:2000ut, Dolan:2003hv, SimmonsDuffin:2012uy,Rychkov} for more details. The motivation to study conformal block is from four point function in CFT, 
\be
\langle \mathcal{O}_1\mathcal{O}_2\mathcal{O}_3\mathcal{O}_4\rangle=(\frac{x_{24}}{x_{14}})^{\frac{\Delta_{ij}}{2}}(\frac{x_{14}}{x_{13}})^{\frac{\Delta_{34}}{2}}\frac{F(u,v)}{x_{12}^{\frac{\Delta_1+\Delta_2}{2}}x_{34}^{\frac{\Delta_3+\Delta_4}{2}}},\label{4pt}
\ee 
where $\Delta_i$ are conformal dimension of operator $\mathcal{O}_i$. $\Delta_{ij}=\Delta_i-\Delta_j$ and $u,v$ are two conformal invariants 
\be
u=\frac{x_{12}x_{34}}{x_{13}x_{24}},\quad v=\frac{x_{14}x_{23}}{x_{13}x_{24}}.
\ee 
The form of four point function \eqref{4pt} is fixed by global conformal invariance up to an unknown function $F(u,v)$ in general. Using OPE of two primary operators, the function $F(u,v)$ can be decomposed into linear superposition of conformal blocks 
\be
F(u,v)=\sum_{\mathcal{O}}C_{12\mathcal{O}}C_{34}^{\mathcal{O}}G_{\Delta,\ell}(u,v)
\ee
which counts the contribution of primary  family of operator $\mathcal{O}$ with dimension $\Delta$ and spin $\ell$. The coefficients $C_{ij}^{\mathcal{O}}$ is three point function constant of operator $\mathcal{O}_i,\mathcal{O}_j$ and $\mathcal{O}$. Any four point function is an eigenfunction of Casimir operator 
\be
\mathcal{L}_2 \langle\mathcal{O}_1\mathcal{O}_2\mathcal{O}_3\mathcal{O}_4\rangle=C_{\Delta,\ell}\langle\mathcal{O}_1\mathcal{O}_2\mathcal{O}_3\mathcal{O}_4 \rangle
\ee 
with eigenvalue
\be
C_{\Delta,\ell}=\Delta(\Delta-d)+\ell(\ell+d-2)
\ee 
in d dimensions. Casimir operator 
\be
\mathcal{L}_2=\frac{1}{2}L_{AB}L^{AB},\quad L_{AB}=L_{AB}^{(1)}+L_{AB}^{(2)}
\ee 
commutes with all generators $L_{AB}$ conformal algebra. Therefore conformal block is an eigenfunction of the following second order partial differential operator 
\be
D_{\upsilon}=x^2(1-x)\frac{\partial^2}{\partial x^2}+z^2(1-z)\frac{\partial^2}{\partial z^2}-(a+b+1)(x^2\frac{\partial}{\partial x}+z^2\frac{\partial}{\partial z})-a b(x+z)+2\upsilon \frac{xz}{x-z}((1-x)\frac{\partial}{\partial x}-(1-z)\frac{\partial}{\partial z}).
\ee 
The parameters $a,b,\upsilon$ are
\be
a=-\frac{\Delta_{12}}{2},\quad b=\frac{\Delta_{34}}{2},\quad \upsilon=\frac{d}{2}-1.
\ee 
\be
D_{\upsilon}G_{\Delta,\ell}(x,z)=\frac{1}{2}C_{\Delta,\ell}G_{\Delta,\ell}(x,z).
\ee 
The variables $x,z$ are related to cross ratios $u,v$ by 
\be
u=x z,\quad v=(1-x)(1-z).
\ee 
The boundary condition is 
\be
G_{\Delta,\ell}(x,z)\sim x^{\frac{\Delta+\ell}{2}}z^{\frac{\Delta-\ell}{2}},\quad z\to0,x\to0.
\ee 
Therefore one could fix the form of conformal block in general dimensions. For example, in four dimensions, 
\bea
G_{\Delta,\ell}(x,z)&=&(-1)^{\ell}\frac{xz}{z-x}(k_{\Delta+\ell}(z)k_{\Delta-\ell-2}(x)-(x\leftrightarrow z)),\nn\\
k_{\beta}(x)&=&x^{\beta/2}{}_2F_1(\frac{\beta-\Delta_{12}}{2},\frac{\beta+\Delta_{34}}{2},\beta,x).
\eea 
In this paper, two spherical regions are at the same time slices,
\be
x=z
\ee  this is the so called ``diagonal limit'' of conformal block \cite{Hogervorst:2013kva}. In diagonal limit, conformal blocks satisfy ordinary differential equations, third order for $\ell=0$ and fourth order for general cases. For equal external operator dimensions, $\Delta_{12}=0$, the closed form of diagonal limit of conformal block is a finite sums of ${}_3F_2$ functions
\bea
G_{\Delta,\ell}(z)=\left\{\begin{aligned}&(1-z)^{-b}F_{\lambda,2n}(Z),\quad &\ell=2n,\\&(\frac{1}{z}-\frac{1}{2})(1-z)^{-b}F_{\lambda,2n+1}(Z),\quad &\ell=2n+1.\end{aligned}\right.
\eea 
The variable $Z$ is 
\be
Z=\frac{z^2}{4(z-1)}.
\ee 
The parameter $\lambda$ is 
\be
\lambda=\frac{\Delta+\ell}{2}.
\ee 
The general form of $F_{\lambda,2n}$ and $F_{\lambda,2n+1}$ are
\bea
\hspace{-10pt}F_{\lambda,2n}(Z)\hspace{-10pt}&=&\hspace{-10pt}(-4Z)^{\lambda-n}\frac{(\lambda-n+\frac{1}{2})_n}{(\lambda-n-b)_n(\lambda-\upsilon-2n+\frac{1}{2})_n}\sum_{r=0}^n \left(\begin{aligned}n\\r\end{aligned}\right)\frac{(b+r)_r(\upsilon+n)_r(\lambda-\upsilon-b-2n)_{n-r}}{(\lambda-n+\frac{1}{2})_r}\nn\\&&\times{}_3F_2(\lambda-n-\upsilon,\lambda-n-b,\lambda-n+b+r;2\lambda-2n-\upsilon,\lambda-n+r+\frac{1}{2};Z),\\
\hspace{-10pt}F_{\lambda,2n+1}(Z)\hspace{-10pt}&=&\hspace{-10pt}(-4Z)^{\lambda-n}\frac{(\lambda-n+\frac{1}{2})_n}{(\lambda-n-b)_n(\lambda-\upsilon-2n-\frac{1}{2})_n}\sum_{r=0}^n \left(\begin{aligned}n\\r\end{aligned}\right)\frac{(b+r)_r(\upsilon+n+1)_r(\lambda-\upsilon-b-2n-1)_{n-r}}{(\lambda-n+\frac{1}{2})_r}\nn\\&&\times{}_3F_2(\lambda-n-\upsilon,\lambda-n-b,\lambda-n+b+r;2\lambda-2n-\upsilon-1,\lambda-n+r+\frac{1}{2};Z).
\eea 
OPE blocks in this work correspond to equal conformal weight  $\Delta_{12}=\Delta_{34}=0$, therefore we just set $b=0$ in this paper. For $d=2$, $\upsilon=0$, one can check the conformal block reduces to  
\be
G_h(\eta)=(-1)^h \eta^h {}_2F_1(h,h,2h,-\eta), \quad \eta=-z,\quad \text{for}\ \Delta=J=h. \label{chiralcb}
\ee 
This is exactly two dimensional conformal block we used in this work. Note for odd $h$, there is an extra minus sign. For general non-chiral operator, $\Delta=h+\bar{h}, J=h-\bar{h}$, the conformal block becomes a product of two \eqref{chiralcb} in the diagonal limit. This is the origin of $\log^2\frac{L}{\epsilon}$ divergence for non-chiral OPE blocks.


\end{document}